\documentclass[sigconf]{acmart}
\usepackage{booktabs} 
\usepackage[utf8]{inputenc}
\usepackage{graphicx}
\usepackage{makecell}

\usepackage[shortlabels]{enumitem}
\usepackage{cases}
\usepackage{xspace}
\usepackage{multirow}
\usepackage{color}
\usepackage{url}
\usepackage[]{caption}
\usepackage[font={}]{subcaption}
\usepackage{amsmath}
\usepackage{amssymb}
\usepackage{amsthm}
\usepackage{bm}
\usepackage{dsfont}
\usepackage{mathtools}

\usepackage{hyperref}
\usepackage{listings}
\usepackage[capitalise]{cleveref}

\creflabelformat{equation}{#2\textup{#1}#3}
\crefname{section}{Section}{Sections}
\Crefname{section}{Section}{Sections}
\crefname{appendix}{Appendix}{Appendices}
\Crefname{appendix}{Appendix}{Appendices}
\crefname{figure}{Figure}{Figures}
\Crefname{figure}{Figure}{Figures}
\crefname{equation}{Equation}{Equations}
\Crefname{equation}{Equation}{Equations}
\crefname{table}{Table}{Tables}
\Crefname{table}{Table}{Tables}
\crefname{mylemma}{Lemma}{Lemmas}
\Crefname{mylemma}{Lemma}{Lemmas}
\crefname{theorem}{Theorem}{Theorems}
\Crefname{theorem}{Theorem}{Theorems}
\crefname{algorithm}{Algorithm}{Algorithms}
\Crefname{algorithm}{Algorithm}{Algorithms}
\crefname{listing}{Query}{Queries}  
\Crefname{listing}{Query}{Queries}
\usepackage{tikz}
\usetikzlibrary{calc,pgfplots.groupplots,arrows.meta}
\usetikzlibrary{decorations.pathreplacing}
\usepackage{pgfplots}
\usepgfplotslibrary{fillbetween}

\newcommand{\verdict}{\textsc{VerdictDB}\xspace}
\newcommand{\verdictshort}{\textsc{Verdict}}
\newcommand{\instacart}{\texttt{insta}\xspace}
\newcommand{\tpch}{\texttt{TPC-H}\xspace}
\newcommand{\synthetic}{\texttt{synthetic}\xspace}
\newcommand{\snappydata}{\textsc{SnappyData}\xspace}

\DeclareMathOperator{\erfc}{erfc}
\DeclarePairedDelimiter\floor{\lfloor}{\rfloor}

\newcommand{\tofixnew}[1]{#1}
\newcommand{\tofix}[1]{#1}

\newcommand{\barzan}[1]{{\color{brown} [B: #1]}}
\newcommand{\yongjoo}[1]{{\color{violet} [Y: #1]}}

\newcommand{\ignore}[1]{}
\newcommand{\ph}[1]{\vspace{1mm} \noindent \textbf{#1}  ---}

\usepackage[normalem]{ulem}
\newcommand{\delete}[1]{{\color{red} \sout{#1}}}

\newcommand{\rev}[2]{#2}

\newcommand{\gs}{\hat{g}}

\def\FigureFolder{figures}

\theoremstyle{theorem}
\newtheorem{theorem}{Theorem}

\newtheorem{definition}{Definition}

\definecolor{darkteal}{HTML}{334960}
\definecolor{darkorange}{HTML}{F46524}
\definecolor{mPurple}{HTML}{9370DB}
\definecolor{mYellow}{HTML}{FFFF99}

\definecolor{tomato}{HTML}{F46524}
\definecolor{seagreen}{HTML}{334960}
\definecolor{seablue}{HTML}{B3B3FF}
\definecolor{darkteal}{HTML}{334960}
\definecolor{darkorange}{HTML}{F46524}
\definecolor{mygray}{HTML}{EBEDEF}
\definecolor{mPurple}{HTML}{9370DB}
\definecolor{mYellow}{HTML}{FFFF99}

\pgfplotsset{
    compat=newest,
    every axis legend/.append style={font=\footnotesize, column sep=5pt},
    /pgfplots/ybar legend/.style={
        /pgfplots/legend image code/.code={
        \draw[##1,/tikz/.cd, bar width=6pt, yshift=-0.25em, bar shift=0pt, xshift=0.8em]
        plot coordinates {(0cm,0.8em)};
        }
    }
}

\pgfplotsset{speedups/.style={
        width=180mm,
        height=32mm,
        ybar,
        ymax=150,
        ymin=0,
        xmin=0.5,
        xmax=33.5,
        bar width=2mm,
        ylabel=Speedup,
        xlabel near ticks,
        ylabel near ticks,
        ylabel style={align=center},
        xtick={1,2,...,33},
        xticklabels={tq-1,tq-3,tq-5,tq-6,tq-7,tq-8,tq-9,tq-10,tq-11,tq-12,tq-13,tq-14,tq-15,tq-16,tq-17,tq-18,tq-19,tq-20, iq-1,iq-2,iq-3,iq-4,iq-5,iq-6,iq-7,iq-8,iq-9,iq-10,iq-11,iq-12,iq-13,iq-14,iq-15},
        xticklabel style={rotate=60,xshift=0mm,yshift=1mm},
        ytick={0, 50, 100, 150},
        yticklabels={0$\times$,50$\times$,100$\times$,150$\times$},
        legend style={
            at={(0,1.1)},anchor=south west,column sep=2pt,
            draw=black,fill=none,font=\footnotesize,line width=.5pt,
            /tikz/every even column/.append style={column sep=40pt}
        },
        legend columns=2,
        every axis/.append style={font=\footnotesize},
        ymajorgrids,
        yminorgrids,
        minor grid style=lightgray,
        nodes near coords,
        every node near coord/.append style={font=\footnotesize,fill=white},
        every node near coord/.append style={rotate=90, anchor=west},
    }}

\setcopyright{acmcopyright}

\acmDOI{10.475/123_4}

\acmISBN{123-4567-24-567/08/06}

\acmConference[SIGMOD'18]{2018 International Conference on Management of Data}{June 10--15, 2018}{Houston, TX, USA}
\acmBooktitle{SIGMOD'18: 2018 International Conference on Management of Data, June 10--15, 2018, Houston, TX, USA}
\acmYear{2018}
\copyrightyear{2018}
\acmArticle{4}
\acmPrice{15.00}

\begin{document}
\title{\verdict: Universalizing Approximate Query Processing}
\titlenote{This is an extended technical report of the paper appearing in SIGMOD'18.}


\author{Yongjoo Park, 
    Barzan Mozafari,
    Joseph Sorenson,
    Junhao Wang}

\affiliation{
  \institution{University of Michigan, Ann Arbor}
  \streetaddress{}
  \city{}
  \state{}
  \postcode{}
}
\email{{pyongjoo,mozafari,jsoren,junhao}@umich.edu}

\begin{abstract}
Despite 25 years of research in academia, approximate query processing (AQP) has had little   industrial adoption.
	One of the major
    causes of
     this slow adoption is the reluctance of traditional vendors
     to make
         radical changes to their legacy codebases, and the preoccupation of 
			newer vendors (e.g., SQL-on-Hadoop products) with implementing standard features. 
Additionally, the few AQP engines that are available are each tied to a specific platform
	and require users to completely abandon their existing databases---an unrealistic 
		expectation given the infancy of the AQP technology.
Therefore, we argue that a universal  solution is needed: 
	a database-agnostic approximation engine
	that will     
		 widen the reach of this emerging technology 
		 	across various platforms.

Our proposal, called  \verdict, uses a middleware architecture that requires no changes to the backend database, and thus,
	can work with all off-the-shelf engines.
Operating at the driver-level, \verdict 
	 intercepts analytical queries issued to the database
	 	and rewrites them into another query that, if 
			executed by any standard relational engine, 
                will yield sufficient information for computing an approximate answer.
\verdict uses the returned result set to compute an approximate answer and error estimates,
	which are then passed on to the user or application.  
However, lack of access to the query execution layer 
	introduces significant challenges in terms of generality, correctness, and efficiency. 
This paper shows how \verdict overcomes these challenges 
	and delivers up to \rev{C6,C9}{171$\times$ speedup (18.45$\times$ on average) for a variety of existing engines, such as Impala, Spark SQL, and Amazon Redshift, while incurring less than 2.6\% relative error.}
\verdict is open-sourced under Apache License.
\end{abstract}

\copyrightyear{2018}
\acmYear{2018}
\setcopyright{acmcopyright}
\acmConference[SIGMOD'18]{2018 International Conference on Management of Data}{June 10--15, 2018}{Houston, TX, USA}
\acmBooktitle{SIGMOD'18: 2018 International Conference on Management of Data, June 10--15, 2018, Houston, TX, USA}
\acmPrice{15.00}
\acmDOI{10.1145/3183713.3196905}
\acmISBN{978-1-4503-4703-7/18/06}



\maketitle




\section{Introduction}
\label{sec:intro}

Despite its long history in academic research~\cite{first-aqp}, approximate query processing (AQP) has had little success in terms of industrial adoption~\cite{mozafari_sigmod2017_invited}.
	Only recently, a few vendors have started to include limited forms of approximation features in their products, e.g.,  
  Facebook's Presto~\cite{presto},
   Infobright's IAQ~\cite{infobright}, Yahoo's Druid~\cite{druid-aqp}, SnappyData~\cite{snappydata_website},
    and Oracle 12C~\cite{oracle-aqp}.
 While there are
	several factors contributing to
 this slow adoption, 
   one of the ways to quickly widen the reach of this technology  
  	is to offer   a \emph{Universal AQP (UAQP)}: 
 	an AQP strategy that could work with all existing platforms without requiring any modifications to existing databases.
\ignore{ one of the key obstacles is 
 	the reluctance of  major vendors in adding new technologies to
		 their legacy codebases.
For example, it took nearly a decade before column stores were adopted by mainstream vendors for analytical workloads. 
However, this reluctance is even more severe for approximate query processing. 
Unlike columnar formats which only affected internal implementation but left the user interface intact (i.e., the familiar SQL semantics), 
	AQP requires modifications to both:  
		incorporate approximation features means changes to both database internals as well as output semantics.
Now, with AQP, users must cope with the uncertainty in their query output.

\delete{While AQP is not suitable to all workloads (e.g., billing), there are numerous applications that are amenable to approximation (e.g.,
	A/B testing, exploratory and interactive analytics, visualization) but cannot use the few AQP engines that are available on the market unless
		they are compatible with their existing platform.
	For example, applications relying on Impala cannot simply use the approximation features offered by SnappyData, unless they first migrate to Apache Spark. }

\ph{Our goal}
An ideal solution to drastically accelerate the adoption of AQP would be a \emph{Universal AQP (UAQP)}: 
	an AQP strategy that could work with all existing platforms without requiring any modifications to existing databases.
 \barzan{Option 2 ends here}}
In this paper, we \ignore{aim to} achieve this goal by
	performing AQP entirely at the driver-level. 
That is,
 we leave the query evaluation logic of the existing database 
	 completely unchanged.
	 Instead, we introduce a middleware that rewrites incoming queries, such that the 
	 standard execution of the rewritten queries under relational semantics would yield
approximate answers to the original queries.
	This requires that the entire AQP process be
 encoded in SQL, including the sample planning, query approximation, and error estimation. 
This approach, therefore, faces several challenges.

\ph{Challenges}
The first challenge is ensuring statistical \emph{correctness}.
When multiple (sample) tables are joined, the AQP engine must account for inter-tuple correlations.
Previous AQP engines have relied on foreign-key constraints~\cite{join_synopses},
	modifying the join algorithm~\cite{wander-join}, or
	modifying the query plan~\cite{kandula2016quickr,mozafari_sigmod2014_diagnosis}. 
However, as a middleware, we can neither change the internal query evaluation nor use non-standard join algorithms. 
With SQL-on-Hadoop systems, we cannot even enforce   foreign-key constraints. 
Thus, we need a different solution that can be implemented by a middleware.
	The second challenge is the \emph{middleware efficiency}.
Pushing the entire computation to the middleware 
	can severely impair performance, because, unlike the database,
		 it is not equipped with query optimization and distributed resources.
Finally, there is a \emph{server efficiency} challenge. 
 For general error estimations, previous AQP engines have resorted to
computationally prohibitive resampling-based techniques~\cite{easy_bound_bootstrap,kleiner2013diagnostic},
intimate integration of the error estimation logic into the scan operator (e.g., \cite{mozafari_sigmod2014_diagnosis,mozafari_cidr2017,kandula2016quickr}),
or even overriding the relational operators altogether~\cite{mozafari_sigmod2014_abm,mozafari_sigmod2014_demo}.
Without access to the query evaluation layer of DBMS, 
	the error estimation has to be expressed as a SQL query,
		which can be extremely expensive and defeat the purpose of approximation.

\ph{Design Criteria}
For our UAQP proposal to be practical, it has to meet three criteria. It must offer sufficient \emph{generality} to support a wide class of analytical queries.
Despite no access to database internals, it must still guarantee \emph{statistical correctness}, i.e., unbiased approximations and error estimates. 
Finally, it must ensure \emph{efficiency}. 
UAQP does not need to be as efficient as a specialized and tightly integrated AQP engine,	
	but to be useful, it still needs to be considerably faster than exact query processing.


\ph{Our Approach}
 First, we sidestep the computational overhead of 
	bootstrap~\cite{easy_bound_bootstrap,mozafari_sigmod2014_diagnosis} and
	the intrusive nature of its analytical variants~\cite{mozafari_sigmod2014_abm}
	by exploiting the theory of \emph{subsampling}~\cite{politis1994large}. 
Note that bootstrap's overhead consists of two parts: the cost of constructing multiple resamples, and the cost of aggregating each resample. The traditional subsampling can only reduce the second part---computational cost---by aggregating smaller resamples. 
Our experiments show, however, the first part---constructing resamples---is still a major performance overhead (\cref{sec:subsampling:efficiency}).


We thus propose a computationally efficient alternative, called \emph{variational subsampling}, 
	which yields provably-equivalent asymptotic properties to traditional subsampling (\cref{thm:exclusive}).
%
	The key observation is that,
instead of running the same aggregation query on different resamples,
	one can achieve the same outcome through a single execution of a carefully rewritten query on the sample table itself. 
The rewritten SQL query treats different resamples separately throughout its execution by relying on a resample-id assigned to each tuple (\cref{sec:subsampling}).
We also generalize this idea to more complex, nested   queries (\cref{sec:subsampling:complex}).
%
%

While integrated AQP engines  use hash tables and counters
	for efficient construction of stratified samples~\cite{kandula2016quickr,mozafari_eurosys2013,surajit-optimized-stratified}, 
	\verdict must rely solely on SQL statements to achieve the same goal.
However, adjusting the sampling probabilities dynamically (according to the strata sizes) while scanning the data 
	can be extremely expensive. 
We thus devise a probabilistic strategy that can be implemented efficiently, by exploiting the properties of a Bernoulli process:
 since the number of tuples sampled per each group follows a binomial distribution,
 	we can (with high probability) guarantee a minimum number of samples per group by adjusting the sampling probabilities accordingly (\cref{sec:stratified}).

Lastly, unlike most AQP engines that use a single sample for each query~\cite{mozafari_sigmod2014_diagnosis,join_synopses,mozafari_eurosys2013,surajit-optimized-stratified,easy_bound_bootstrap} (or generate samples on the fly~\cite{kandula2016quickr}),
	\verdict can choose and combine multiple samples that minimize error (among those prepared offline), given an I/O budget.



\ph{Contributions} We make several contributions:
\begin{enumerate}[1.,nolistsep,noitemsep,leftmargin=10pt]
\item We explore the idea of a Universal AQP as a platform-agnostic solution that 
  can
 work with any database system without any modifications. 
	We propose a realization of this idea, called \verdict, 
		that relies solely on a middleware architecture and implements the entire AQP process via re-writing SQL queries.
		
\item We develop an efficient strategy for constructing stratified samples that 
	provide probabilistic guarantees (\cref{sec:stratified}). We also propose a novel technique, called \emph{variational subsampling},
		which enables faster error estimation for a wide class of  queries and 
			can be efficiently implemented as SQL statements (\cref{sec:subsampling,sec:subsampling:complex}).
	


\item We conduct extensive experiments on both benchmark and real-world sales datasets
	 using several modern query processors (Impala,  Redshift, and Spark SQL).
	Our results show that \verdict speeds up these database engines on average by 57$\times$ (and up to 841$\times$), while incurring
		less than 2.6\% error.
\end{enumerate}
\rev{C3}{While a few other AQP systems have also relied on query-rewriting \cite{aqua2,join_synopses,galakatos2017revisiting},
    our techniques enable a wider range of practical AQP queries on modern SQL-on-Hadoop engines (see \cref{sec:related} for a detailed comparison).}

\vspace{2mm}

\noindent

The remainder of this paper is organized as follows. \cref{sec:overview} provides a high-level overview of \verdict. 
\cref{sec:preparation} describes \verdict's sample preparation and its novel technique for stratified sampling.
 \cref{sec:subsampling} explains the basics of our error estimation technique, which are then extended to joins and nested queries in \cref{sec:subsampling:complex}.
	\cref{sec:experiments} presents our experiments, followed by an overview of the related work in \cref{sec:related}.
	Finally, \cref{sec:conclusion} discusses our future plans.
	


\ignore{
\yongjoo{END OF INTRO}
			
In recent years, the database community has paid a significant amount of effort for the development of sophisticated AQP techniques~\cite{park2017active,mozafari_pvldb2015_ksh,mozafari_icde2016,crotty2015vizdom,haas:large-sample,aqua1,join_synopses,dynamicp-sample-selection,surajit-optimized-stratified}. In general, those techniques could bring 100$\times$--200$\times$ speedups by approximately answering common analytic queries based on offline-created samples. Despite this technical advance and large performance benefit, we barely find any actual database systems---either open-sourced or commercialized---that have adopted those AQP techniques.

We observe that this slow adoption originates from an implicit assumption commonly made by previous AQP techniques; that is, for maximum performance benefit, they assumed that can significantly modify the query evaluation logic within existing relational database systems~\cite{mozafari_eurosys2013,kandula2016quickr,mozafari_sigmod2014_diagnosis}. However, existing database systems with well-maintained codebases do not incorporate such drastic changes in exchange for stability. Also, database administrators are reluctant to employ another AQP-dedicated cluster especially when they have a large amount of corporate data managed by an existing database system with complicated authentication and authorization mechanisms.

In this work, we close this gap, for the first time in the literature, by developing a suite of AQP techniques \emph{that are completely independent from the relational query evaluation logic within database systems.} In other words, an AQP engine, that employs the techniques presented in this work, can process queries by communicating with existing database systems solely relying on SQL; however, the AQP-nature of the engine enables hundreds-times faster query processing. Moreover, our techniques presented in this paper are \emph{complete}; our techniques cover all operations necessary for a standalone AQP solution, including sample preparation, query planning, and error estimation. This implies that the AQP engines employing our techniques can provide interactive data analytics experience as running on top of any existing SQL-supporting query processor, such as Apache Spark\cite{sparksql}, Apache Impala~\cite{impala}, Apache Hive~\cite{hive}, Amazon Redshift~\cite{redshift}, Google BigQuery~\cite{bigquery}, and so on. We call this approach \emph{Universal AQP (UAQP)}.

\ph{Challenges}
\tofix{
The primary challenge of AQP is providing statistically correct error bounds for its approximate answers in a computationally efficient way for a wide class of analytic queries. The additional challenge of Universal AQP is developing efficient, \emph{SQL-based} algorithms for error bound computations without compromising the generality of the supported queries.}

\tofix{\Cref{tab:err_est} summarizes the properties of known error estimation methods. The table tells us that there exist no efficient SQL-based error estimation methods that can support the queries including non-FK-joined tables. Note that HDFS-based distributed databases, such as Hive, Impala, Spark SQL, do not have FK constraints; thus, enforcing referential integrity is unavailable in the first place.}




\ph{Design Criteria}
Successful UAQP should satisfy the following three design criteria. \tofix{First, UAQP must be computationally efficient to bring sufficient speedups compared to the systems without AQP. Second, UAQP should provide unbiased estimates accompanied with statistically correct error bounds for the queries including multiple tables joined through non-FK relationships. Third, UAQP should support complex queries that include joins, derived tables, subqueries in selection predicates, etc.}

\begin{table}[t]
\centering
\small
\begin{tabular}{p{32mm} p{27mm} p{11mm} }
\Xhline{2\arrayrulewidth}
\textbf{Method} & \textbf{Correct error bounds}\newline \textbf{for non-FK joins}
 & \textbf{Efficient}\newline \textbf{in SQL} \\ \hline
Simple mean with CLT~\cite{mozafari_eurosys2013} & No & Yes \\
Horvitz-Thompson~\cite{kandula2016quickr} & Yes & No \\
Bootstrap~\cite{easy_bound_bootstrap,mozafari_sigmod2014_diagnosis} & Yes & No \\
Na\"ive subsampling & Yes & No \\
\textbf{Ours (some name)} & Yes & Yes \\ \Xhline{2\arrayrulewidth}
\end{tabular}
\caption{Comparison of error estimation methods.}
\label{tab:err_est}
\end{table}

\ph{Our Approaches}
Our UAQP techniques take the following approaches to satisfy the three design criteria mentioned above.

\tofix{First, we employ an approach, called \emph{subsampling}~\cite{??}, that can be $O(b)$ times more efficient, when I/O cost is ignored, than bootstrap. Subsampling can also, similar to bootstrap, estimate the expected errors (in terms of variance) of any smooth aggregate functions based on the statistics of resamples. The crucial difference is that the sizes of those resamples can be smaller than $n$.}
\tofix{However, a na\"ive application of subsampling to query processing is inefficient since it needs to scan an entire sample for constructing every resample anyway. Our technique simulates subsampling in a way that requires only a single scan over the data.}


\tofix{Second, we show that our variant of subsampling computes statistically correct variance of sample statistics; thus, our approach provides statistically correct error bounds.}

\tofix{Third, we show that our variant of subsampling can be extended to support complex queries including nested subqueries, joins of multiple tables, etc.}

\tofix{In addition, we present an alternative stratified sampling process that can be efficiently implemented in SQL. Our stratified sample process does not rely on counters or hash tables needed to provide absolute guarantees on no missing groups; instead, our technique provides the probabilistic guarantee that no groups are missed with probability at least $1-\delta$. Also, unlike the previous work~\cite{??} which finds only a single sample table, our techniques finds an optimal set of sample tables that could produce most accurate answers.}




We have implemented all our UAQP techniques in an open-source query processing system called \verdict~\cite{verdict}. Currently, \verdict can run on top of Apache Hive, Apache Impala, Apache Spark SQL (both 1.6 and 2.0), and Amazon Redshift.

}


\section{System Overview}
\label{sec:overview}

In this section, we provide a high-level overview of \verdict's components and operations. 
In \cref{sec:architecture}, we briefly introduce \verdict's deployment architecture and its internal components. 
In \cref{sec:queries}, we discuss the types of SQL queries that are sped up by \verdict. 
Lastly, in \cref{sec:workflow,sec:interface}, we explain \verdict's query processing workflow and its user interface. 


\subsection{Architecture}
\label{sec:architecture}

We first describe how \verdict works with other parties (i.e., users and a database system). Then, we describe the internal components of \verdict.

\input{\FigureFolder/architecture}

\ph{Deployment Architecture}
As depicted in \cref{fig:deployment},
\verdict is placed between and interacts with the \emph{user} and an off-the-shelf database.
  We call the database used alongside \verdict the \emph{underlying database}.
  The user can be a data analyst who issues queries through an interactive SQL shell or visualization tool, or any application that issues SQL queries. 
	The user sends queries to \verdict and obtains the query result directly from \verdict without interacting with the underlying database.

\verdict communicates with the underlying database via SQL for obtaining metadata (e.g., catalog information, table definitions) 
	and for accessing and processing data. 
For this communication, \verdict  uses the standard interface supported by the underlying database, such as JDBC for Hive and Impala,
	ODBC for SQL Azure,
	or \texttt{SQLContext.sql()} for Spark.\footnote{\texttt{SparkSession.sql()} for Spark 2.0 and above.} 
Note that the contributions presented in this work are applicable irrespective of these specific interfaces and can be applied to any 
	database that provides an interface through which SQL statements can be issued.
\rev{C10}{\verdict requires the underlying database to support \texttt{rand()}, a hash function (e.g., \texttt{md5}, \texttt{crc32}), window functions (e.g., \texttt{count(*) over ()}), and \texttt{create table ... as select ...}.}

\verdict stores all its data, including the generated samples and the necessary metadata, in the underlying database. 
\ignore{Storing the constructed sample tables in the underlying database provides the following three benefits. First, Verdict can provide speedups even when a user connects to Verdict on a client machine different from the one on which the samples were created. Second, once sample tables are created, all other users also benefit. Third, the underlying database can easily access and process rewritten queries including both sample tables and original tables, which typically happens when small dimension tables are included in queries.} 
	\verdict accesses the underlying database on behalf of the user (i.e.,using his/her credentials);
	thus, \verdict's data access privilege naturally inherits the data access privileges granted to its user. 

\ph{Internal Architecture} \Cref{fig:internal_architecture} shows  \verdict's  internal components. Given a SQL query, Query Parser translates it 
	into logical operators (e.g., projections, selections, joins, etc.).
 Then, AQP Rewriter converts this logical expression into another logical expression that performs AQP (\cref{sec:subsampling,sec:preparation}).

  Syntax Changer
 	 converts this rewritten logical expression into a  SQL statement that can be executed on the underlying database. 
	This is the only module in \verdict that needs to be aware
	of the DB-specific limitations
	 (e.g., no \texttt{rand()} permitted in selection predicates in Impala)
	  and its SQL dialects (e.g., quotation marks, different function syntaxes for \texttt{mod}, \texttt{substr}, etc.).
 This allows \verdict to easily support new databases.\footnote{\verdict's current release comes with drivers for  
	Apache Hive, Apache Impala, Apache Spark SQL (1.6 and 2.0), and Amazon Redshift. We plan to add drivers for Oracle, Presto, and HP Vertica in the near future.}
To add support for a new DBMS, 
	the only part that needs to be added to \verdict is a thin driver that extends that DBMS's JDBC/ODBC driver and 
		understands its SQL dialect.
\verdict's implementation is 57K lines of code (LOC), while adding a driver for
Impala, Spark SQL, and Redshift required only
55, 167, and 360 LOC, respectively.  

Once the rewritten query is executed by the underlying database,   Answer Rewriter adjusts the results  (e.g., output format, error reporting format, confidence levels, etc.) and
returns
 an approximate answer (and error estimates, when requested) to the original query.

\begin{table}[t]
\centering
\small
\begin{tabular}{p{17mm} p{55mm}}
\hline
\textbf{aggregates} & \texttt{count}, \texttt{count-distinct}, \texttt{sum}, \texttt{avg}, \texttt{quantile}, user-defined aggregate (UDA) functions
 \\ \hline
\textbf{table sources} & derived tables or base tables joined via equi-joins;
the derived table can be a select statement with or without aggregate functions.
 \\ \hline
\textbf{selections} \newline \textbf{(filtering)} &
 expr comp expr (e.g., \texttt{price > 100}), \newline
 expr comp subquery (e.g., \texttt{price > (select ...)}),
 logical \texttt{AND} and \texttt{OR}, etc. \\ \hline
\textbf{other clauses} & \texttt{group by}, \texttt{order by}, \texttt{limit}, 
 \texttt{having} \\ \hline
\end{tabular}

\vspace{2mm}

\caption{Types of queries that benefit from \verdict.}
\vspace{-6mm}
\label{fig:supported_queries}
\end{table}

\subsection{Supported Queries}
\label{sec:queries}

\verdict speeds up analytic SQL queries that use common aggregate functions. 
When \verdict can speed up a query, we say \verdict \emph{supports} that query. 
Other queries are simply passed down to the underlying database unchanged, i.e., unsupported queries do not observe any speedup. 
Currently, \verdict supports queries
	with \emph{mean-like} statistics, 
		including common aggregate functions (e.g., \texttt{count}, \texttt{sum}, \texttt{avg}, \texttt{quantile}, 
		\texttt{var}, \texttt{stddev}),
			 	and user-defined aggregates (as long as they converge to a non-degenerate distribution~\cite{politis1994large}).
\verdict supports \texttt{count-distinct} using a function that partitions a domain into subdomains with equal cardinalities~\cite{flajolet2007hyperloglog}.
\verdict does not approximate extreme statistics (i.e., \emph{min} and \emph{max}). 
	Although there is theoretical work on estimating extreme statistics~\cite{smith1990extreme}, 
		the error bounds tend to be quite large in practice. 
However, if a query includes both extreme statistics and other mean-like statistics, \verdict automatically decomposes the query into one part with extreme statistics and the other part with mean-like statistics; then, it approximately computes only the part with mean-like statistics.

\verdict also supports equi-joins, comparison subqueries (e.g., \texttt{where sales < (select avg(sales) ...)}), and other selection predicates (e.g., \texttt{IN list}, \texttt{LIKE regex}, \texttt{<, >}, and so on).
When there is a comparison subquery,
 \verdict  converts it into a join.
  For instance, consider the following query with a correlated subquery:

\begin{lstlisting}[
    basicstyle=\footnotesize\ttfamily,
    xleftmargin=10pt
]
select ...
from orders t1 inner join order_products t2
  on t1.order_id = t2.order_id
where price > (select avg(price)
                 from order_products
                where product = t1.product);
\end{lstlisting}
This query produces the same results as the following query, which instead uses a join with a derived table.

\begin{lstlisting}[
    basicstyle=\footnotesize\ttfamily,
    xleftmargin=10pt
]
select ...
from orders t1 inner join order_products t2
  on t1.order_id = t2.order_id
  inner join (select product, avg(price) avg_price
              from order_products
              group by product) t3
  on t2.product = t3.product
where t2.price > avg_price;
\end{lstlisting}
\tofix{The above query flattening is performed for comparison subqueries.
Currently, \verdict does not approximate
  other types of subqueries, e.g., \texttt{IN (select ...)}, \texttt{EXISTS (select ...)}, or subqueries in the select clause.}
\Cref{fig:supported_queries} summarizes the types of queries  supported by \verdict.

\subsection{Workflow}
\label{sec:workflow}

The user's interaction with \verdict consists of two stages: sample preparation and query processing. 
The sample preparation stage is an \emph{offline} process, during which the user informs \verdict of the tables for which AQP is desired. 
By default, \verdict automatically builds different types of sample tables
  based on the column cardinalities (see~\cref{sec:default_sampling}); 
  however, the user can also manually specify which types of sample tables to build. 
In general, \verdict may construct multiple (sizes and types of) samples for the same table.
 \Cref{sec:preparation} describes the different types of sample tables that \verdict constructs. 
 The metadata about the created sample tables (e.g., names, types, sampling ratios) are recorded in a
 	specific schema inside the database catalog.

\input{\FigureFolder/fig_workflow}

At runtime, when the user issues a query, \verdict first identifies the set of sample tables that can be used in place of each of the base tables that appear in the query. 
Then, \verdict's sample planning module determines a combination of sample tables that can minimize the overall approximation error 
	given a specified I/O budget (e.g., 2\% of the original data).
Depending on the available samples, the I/O budget, or the query type,
	the sampling module may simply resort to using the base tables themselves \cref{sec:planning}.

Once a combination of sample tables is chosen to use for query processing, \verdict rewrites the original query into another SQL statement that,
	when executed by the underlying database, 
	can simultaneously produce both an unbiased approximate answer and probabilistic error bounds.
	When  the underlying database returns the result for the rewritten query,  
		\verdict extracts
			and scales the approximate answer and the error estimates, 
				and returns to the user.
This workflow is visualized in \Cref{fig:workflow}.

For simplicity, we present our techniques assuming that the data in the original tables are static.
\verdict can also efficiently support periodic data ingestion
\cref{sec:appends}.



\subsection{User Interface}
\label{sec:interface}



Traditionally, AQP engines have allowed users to either specify 
 a latency requirement (e.g., return an approximate answer within 2 seconds)~\cite{mozafari_pvldb2012,mozafari_eurosys2013}, or 
 	an accuracy requirement (e.g., return an answer that is at least 99\% accurate)~\cite{mozafari_eurosys2013,mozafari_pvldb2012,kandula2016quickr,nirkhiwale2013sampling}.
The problem with offering a latency knob is that predicting the latency of a query in advance, even when the input size is known, is still
	an unsolved problem for databases~\cite{leis2015good,mozafari_cidr2013,mozafari_sigmod2013}.\footnote{It is well-known that the cost estimates provided by query optimizers 
		 are not an accurate predictor of actual latencies.}
For example, previous engines offering latency knobs have resorted to simple heuristics (e.g., linear regression in BlinkDB~\cite{mozafari_eurosys2013}),
	which often miscalculate    actual runtimes.
Likewise, predicting the approximation error before running the query is practically impossible~\cite{approx_chapter}.
Even when closed-form error estimation is applicable~\cite{kandula2016quickr}, the estimate depends on several query-dependent factors, such as the query selectivity, 
the variance of the attribute values satisfying the selection predicates, the inter-tuple correlations (when joining sample tables), etc. 
	These query-dependent factors are hard to predict, and are typically known only after the AQP engine has run the query.

\ignore{
With the exception of trivial queries (whose error has a closed-form),
 \barzan{this is not true! QuickR doesn't use bootstrap and claims everything. plz
read their HT estimators. we need to explain why their technique is not implementable as a middleware or else this single objection
will kill ur only contribution which is subsampling}
	the error estimation, in general, requires the query output on many small samples 
			in order to construct an empirical error distribution (e.g., bootstrap~\cite{g-ola,zeng2016iolap,mozafari_sigmod2014_abm}).
In other words, the AQP engines cannot predict whether a particular sample  
		will satisfy the user's accuracy requirement, until they have run the query on that sample. }
		
For these reasons, \verdict offers a more practical knob to the user, which is easier to enforce. 	
	Instead of specifying a latency or accuracy requirement, 
		\verdict's users specify an I/O budget.
For every table that exceeds a  certain size (10 million rows, by default), 
	users can choose a maximum percentage of the table that can be used 
		when that table appears in analytical queries (2\%, by default).
Optionally, users can also specify a minimum accuracy requirement. 
	However, \verdict interprets  this accuracy requirement only after the query is executed  and the approximation errors are estimated:
 	if the error(s) violate the accuracy requirement,
		\verdict  reruns the query on the base tables themselves and returns an exact answer back to the user.
In such cases, \verdict uses the notion of High-level Accuracy Contract (HAC) \cite{approx_chapter}, which is
also adopted by SnappyData~\cite{mozafari_cidr2017,mozafari_sigmod2016_demo}.
Similar to previous AQP engines~\cite{olteanu2010approximate,canty2006diagnostics,online-agg-mr1,easy_bound_bootstrap,earl,kleiner2013diagnostic,aqua2,join_synopses,random-sampling-joins}, 
	the error semantics in \verdict are based on the notion of confidence intervals.
For instance, 99\% accuracy at 95\% confidence would mean  that the true answer lies between $\pm 1\%$  of the approximate answer 
	with 95\% 	
	probability.

Similar to Oracle 12c~\cite{oracle-aqp},
  the approximation settings in \verdict (e.g., I/O budget, accuracy requirements)
  		can be set either on a per-query basis or at the system/connection level.
The latter allows \verdict to be used in a \emph{transparent} mode
for speeding up (legacy) applications that are not designed to
	use AQP features, e.g., the DBA can choose appropriate settings on behalf of the application.
\verdict does not include error estimates as additional columns in the output, unless 
	requested by the user.
Again, this is to  ensure that legacy applications can seamlessly consume approximate results 
  without the need to be modified.
		
%
%


\section{Sample Preparation}
\label{sec:preparation}

In
\Cref{sec:preparation:overview}, we briefly review the four types of sample tables
used by
\verdict: uniform samples,
    hashed samples, stratified samples, and irregular samples.


With the exception of stratified samples, the rest of the sample types can be constructed using SQL in a straightforward manner.
For stratified samples, the sampling ratios  differ for each stratum;
however, adjusting the sampling probabilities dynamically while scanning the data using procedural SQL (e.g., Transact-SQL) is not applicable
	 to all databases, and can also be much slower than standard select statements.
 In \cref{sec:stratified}, we introduce \verdict's probabilistic approach to efficient construction of stratified samples,
 	which relies only on standard select statements.

%

\subsection{Background: Sample Types}
\label{sec:preparation:overview}

A sample table $T_s$ is a subset of tuples from an original table $T$. The inclusion probabilities (a.k.a. sampling probabilities)
	 may differ for each tuple. In addition to the tuples themselves, \verdict also records their sampling probabilities
		as an extra column in the sample table.
We can define different sample types using
 a real-valued sampling parameter $\tau\in [0,1]$, 
a column set $\mathcal{C}$ (e.g., $\mathcal{C}$ $=$ $\langle$\texttt{city}, \texttt{age}$\rangle$),
 and the number of unique values $d_C$ under $\mathcal{C}$.
\begin{enumerate}[1.,nolistsep,noitemsep,leftmargin=10pt]
\item \textbf{Uniform sample.} A sample $T_s$ is uniform if every tuple in $T$ is sampled independently (i.e, a Bernoulli process) with a sampling probability equal to $\tau$.

\item \textbf{Hashed sample.}\footnote{Hashed samples are also called \emph{universe samples}~\cite{kandula2016quickr,hadjieleftheriou2008hashed}.}
Given a column set $\mathcal{C}$,
	a hashed sample on $\mathcal{C}$ is defined as $T_s = \{t\in T \mid h(t.\mathcal{C}) < \tau\}$,
		where $h(\cdot)$ is a uniform hash function
         that maps every value of $\mathcal{C}$ into a real number in $[0,1]$,
	and  $t.C$ is the value of $\mathcal{C}$ in $t$.
Here, the sampling probabilities are all set to $|T_s| / |T|$.
\ignore{should sampling prob be
0 or 1 in this case or |S|/|D| for everyone? i am starting to think it should be the latter actually
I'd like to choose this, because this is more easily understandable. Technically, there should be two types of sampling probabilities, i.e., sampling ratio on $\mathcal{C}$ and sampling ratio within $\mathcal{C}$. Explaining this is not a contribution, but will take lots of space.}

\item \textbf{Stratified sample.}
Given a column set $\mathcal{C}$ with unique values
$\{c_1, \ldots, c_{d_{\mathcal{C}}}\}$,
	 a stratified sample on $\mathcal{C}$ is a sample that satisfies this condition:
 \begin{align}
\forall i = 1, \ldots, d_{\mathcal{C}}:
\quad
\left| \sigma_{\mathcal{C}=c_i}(T_S) \right|
 \ge \min \left(
           \frac{|T| \cdot \tau}{d}, \,
          \left| \sigma_{\mathcal{C}=c_i}(T) \right|
        \right)
\label{eq:stratified}
\end{align}
The sampling probability of a tuple with $c_i$ in $\mathcal{C}$ is set as
$\frac{|\sigma_{\mathcal{C}=c_i}(T_S)| }{ |\sigma_{\mathcal{C}=c_i}(T)|}$.


\item \textbf{Irregular sample.}
 When the sampling probabilities do not meet any of the properties mentioned above,
    we call it an irregular sample.
\end{enumerate}

\vspace{2mm}

\noindent During the sample preparation stage, \verdict only constructs sample tables that belong to one of the first three types, i.e., uniform sample, hashed sample, and stratified sample. Irregular samples may  appear only during query processing as a result of joining
    other (sample) tables. 
%
%
%
%
By default, \verdict uses 1\% for $\tau$ so that
the sample sizes
 are within the default query-time I/O budget (i.e., 2\%).


\subsection{Probabilistic Stratified Samples}
\label{sec:stratified}

This section presents \verdict's SQL-based, easily parallelizable approach to constructing stratified samples.
\verdict takes a two-pass approach for creating stratified samples:
in the first pass, the group sizes are computed; in the second pass, tuples are sampled according to group-size-dependent sampling probabilities, as follows.

\begin{lstlisting}[
    basicstyle=\footnotesize\ttfamily,
    xleftmargin=5pt
]
select *
from orders inner join T_temp
     on T.c1 = T_temp.c1 and ... and
        T.ck = T_temp.ck
where rand() <
  (sampling_prob_expression)
\end{lstlisting}
where \texttt{T\_temp} is the table constructed in the first pass with the schema $\langle \texttt{c1}, ..., \texttt{ck}, \texttt{strata\_size} \rangle$.
Here, $\langle \texttt{c1}, ..., \texttt{ck} \rangle$ is the column set of the stratified sample,
 and \texttt{(sampling\_prob\_expression)}
 is a SQL expression that determines the sampling probability for each tuple, which we describe in detail below.

Note that the tuples here are sampled independently from one another (i.e., Bernoulli process). The key advantages of this approach are that (1) the sampling process can easily be expressed in SQL, and (2) its operations can be executed in parallel.

However, the downside here is that the guarantee in \cref{eq:stratified} may no longer hold.
 This is because a Bernoulli process does not produce a sample with exactly $p\%$ of the tuples. For example, suppose that we need to sample at least 10 tuples out of a stratum of 100 tuples (i.e. \texttt{strata\_size} = 100).
If we use a Bernoulli process with a sampling ratio of
     0.1 (10 / 100), we will have fewer than 10 tuples with probability
  $\sum_{k=0}^9 \binom{100}{k} \, 0.1^k \, 0.9^{100-k} \approx 0.45$.
  In other words,  a na\"ive approach would violate the  guarantee of \cref{eq:stratified} for nearly half of the strata.

To guarantee \cref{eq:stratified}, \verdict uses a staircase function by substituting  \texttt{(sampling} \texttt{\_prob\_expression)} with a case expression, i.e., \texttt{(case strata\_size > 2000 then 0.01 when strata\_size > 1900 then 0.012 ... else 1)}. The staircase function expressed in a case expression upper-bounds $f_m(n)$, where $f_m(n)$ is a value such that
  a Bernoulli process with ratio $f_m(n)$ samples at least $m$ out of $n$ tuples with probability $1-\delta$ (by default, $\delta$=0.001).
\verdict uses the following lemma to determine $f_m(n)$ (proof deferred to \cref{sec:proofs}).

\ignore{\verdict computes the sampling ratio relying on the properties of Bernoulli sampling. Specifically, we derive a formula for a sampling ratio such that it can guarantee the minimum group size with predefined (high) probability (i.e., $1 - \delta$). Here, $\delta$ is a small number (e.g, 0.01) that indicates the chance in which the guarantee is violated. The following theorem provides precise sampling ratios.}

\begin{lemma}
Let a sample be constructed by Bernoulli sampling from $n$ tuples with $p$ sampling probability.
Then, the sampling probability for outputting at least $m$ tuples with probability $1 - \delta$ is
\begin{align*}
f_m(n) &= g^{-1}(m; n) \\
\text{where} \quad
g(p; n) &= \sqrt{2 n \cdot p (1 - p)} \, \erfc^{-1} \left( 2 (1 - \delta) \right) + n \, p
\end{align*}
$\erfc^{-1}$ is the inverse of the (standard) complementary error function.
\label{thm:stratified}
\end{lemma}



\ignore{
\yongjoo{Old material with comments in place.}

In
\Cref{sec:preparation:overview}, we briefly review the four types of sample tables
used by
\verdict: simple random samples (a.k.a. uniform samples),
	hashed samples, stratified samples, and irregular samples.
As mentioned in \cref{sec:workflow}, \verdict constructs certain samples by default (details in \cref{sec:default_sampling}) but users can also manually
	request certain types of samples to be built.
%
%
\barzan{added this, see if its correct}
These sample types can be constructed using SQL in a straightforward manner, with the exception of stratified samples:
\barzan{here say why u re singling out strat samples and u only have an efficient alg for them?}
%
 In \cref{sec:stratified}, we introduce \verdict probabilistic approach for efficient construction of stratified samples.
	using SQL. 

\subsection{Background: Sample Types}
\label{sec:preparation:overview}

As previously mentioned,
various types of sample tables in \verdict are constructed during the offline sample preparation stage, which are used at query time for AQP.
Let $T$ be a table with $N_T$ tuples. \barzan{i'd get rid of $_T$ subscript if u dont need to talk about multiple tables to keep it simple!.
Also, I didn't see N used in \cref{sec:preparation:overview} at all. You can remove it and add define it where u need it}
Sampling is a random process $f$  selects a subset $S$ of the tuples from $T$ according to some probabilities stored in a vector
$\vec{p}\in [0,1]^{N_T}$, i.e., each element of $\vec{p}$ is a real value in $[0,1]$.
\verdict uses a
Bernoulli process to generate a sample table: it considers each tuple in $T$ independently, and according to its corresponding value in $\vec{p}$.
We use $S$ to refer to the specific sample that is generated by this process.
\barzan{I thought  $f$ was unnecessary and also the distinction between a random table and its instance. bring back if u think otherwise}


Let $\mathcal{C}$ be a subset of columns in $T$, and $|\mathcal{C}|$ \barzan{try to change this notation. cardinality means how many cols not number of unique vals}
denote the number of unique values of $C$ that appear in $T$.
\barzan{i got rid of `domain' since a domain size can be much larger than the number of actual vals that appear in a column set.
see if u use `domain' anywhere else in the paper}
  For instance, if $\mathcal{C}$ is \{\texttt{city, age}\} and takes on the following values \{NYC, 30\}, \{NYC, 31\}, \{LA, 31\} in $T$,
  	\barzan{i think using \{.\} is a bit awkward for this, unless u have used it before. u can instead say $\{\langle A,B\rangle, \langle C, D\rangle\}$ or just not have
	\{.\} at all. see what's used elsewhere and be consistent}
  	then |\{\texttt{city, age}\}| is three. \barzan{changed this example  see if its what u meant}
We can  formally define each sample type according to the properties of the vector $\vec{p}$,
	which is defined based on a given \emph{sampling rate} $0<\tau\leq 1$.

\begin{enumerate}[1.,nolistsep,noitemsep,leftmargin=10pt]
\item \textbf{Simple random sample.} In this type of sample, all the values in $\vec{p}$ are equal to $\tau$,
	i.e., the same sampling probability for all tuples.

\item \textbf{$\mathcal{C}$-hashing sample.}\footnote{Hashing samples are also called \emph{universe samples}~\cite{quickR}\barzan{see if u can
find other names or references for this}.} Given parameter $\tau$ and a subset of columns
 $\mathcal{C}$, the probabilities in $\vec{p}$ for a  $\mathcal{C}$-hashing sample
  are either 0 or 1,  depending on the value of $\mathcal{C}$ in each tuple.
 For a tuple whose value of columns $C$ is $c$,  the corresponding probability in $\vec{p}$
 	is 1 when $h(c) < \tau$ and is 0 otherwise.
Here, let $h(.)$ be a \barzan{what is the property of this hash function for this to work?} hash function that maps every value of
	 columns $C$ to a real value between 0 and 1.
Note that different tuples with the same values for $\mathcal{C}$ are either all included in the sample (probability 1) or
	are all excluded (probability 0).

\item \textbf{$\mathcal{C}$-stratified sample.}
\barzan{Give the formal definition here since u ve already bored the heck out of the reader w/ all this notation!}
\tofix{ Let $S_c$ be a set of the tuples that have $c$ as attribute values in the column set $\mathcal{C}$. For example, if $c$ is \{``New York City''\} where $\mathcal{C}$ is \{ \texttt{city} \}, then $|S_{\{``New York City''\}}|$ is the number of the tuples that have \{``New York City''\} in the \{\texttt{city}\} column set. Informally, the sampling probabilities for $\mathcal{C}$-stratified sample are assigned so that $|S_c|$ (i.e., groups sizes in a sample table) are as balanced as possible. The sampling probabilities are determined using a function of $|T_c|$, where $T_c$ is the set of the tuples in $T$ that have $c$ in $\mathcal{C}$.
The function of $|T_c|$ is described in \cref{sec:stratified}. In general, the smaller $|T_c|$ is, the larger its sampling probability is.}

\item \textbf{Irregular sample.} \barzan{changed irregular random sample to irregular sample. make consistent everywhere}
 When the sampling probabilities do not meet any of properties mentioned above,
	we call it an irregular sample.
\end{enumerate}


In the sample preparation stage, \verdict only constructs sample tables that belong to one of the first three types, i.e., simple random sample, $\mathcal{C}$-hashing sample, and $\mathcal{C}$-stratified sample. Irregular random samples may  appear only during query processing as a result of joining
	other (sample) tables (\cref{sec:planning}).
For  sample tables that are constructed offline, \verdict records their types and sampling parameters (e.g., $\tau$ or $\mathcal{C}$)
	in its metadata tables.
\tofix{To store $\vec{p}$, \verdict augments a sample table with an extra column.}
\barzan{this is strange since u earliler defined $\vec{p}$ to have the same length as the size of the original table!}

\subsection{Stratified Sample Construction}
\label{sec:stratified}

\barzan{say something about how strat samples are built in 1) BlinkDB  and 2) QuickR (read QuickR paper's strat strategy very carefully since it's different than Blink and is one-pass sampling!).
Then explain why both approaches are inefficient for us as a middleware, e.g., explain how they require access to internals etc.}

\verdict therefore takes a different approach to constructing its stratified samples.
Instead of exactly computing the probabilities from equation (\ref{eq:??}),
	\verdict constructs a probabilistic stratified sample. In other words,
	\barzan{here explain the role of delta and the new equation or its semantic or how it changes that equation from 3.1}
\verdict constructs these probabilistic stratified samples
	in two steps. First, it computes the sampling probabilities.
	Second, it composes a SQL statement that uses those probabilities to construct the stratified sample.

\ph{1. Computing the Probabilities}
\barzan{i feel like the following is too wordy and should be moved to 3.1. Here, you should just give a reference to the equation number}
\tofix{\verdict's stratified sample creation pursues the following two goals. First, the group size in a sample table, i.e., $|S_{c}|$, should be as balanced as possible. Second, at least $m$ number of tuples must be sampled for every group, where $m$ is a user-configurable parameter.}
Given a column set $\mathcal{C}$ and an I/O budget \tofix{blah}, \barzan{1) try to not use m as u already defined tau in 3.1. 2) u need to explain
how to compute tau or m from the user specified I/O budget}

To compute the sampling probabilities using equation (\ref{eq:??}),  \verdict
	has to compute two quantities for every tuple in $T$: \barzan{add something here based on what ur equation shows in sec 3.1}
 However, these sampling probabilities are the same for tuples  that share the same values for the columns  in $\mathcal{C}$.
 \barzan{is this prev sentence correct even for verdict? or is it only true for non-probabilistic strat sample?}

The first quantity can be easily computed as $N_T / |\mathcal{C}| \times \tau / |T_{c}|$, \barzan{explain why if it's not exactly what's in the formula from 3.1}
 \barzan{these notations are terrible: do NOT use |.| to mean number of unique vals. also, what the heck is $T_c$?? never defined it. just use projection symbol
 from relational algebra if u can}
which requires \barzan{explain what query u need to run to get these
	variables and why it is cheap to do so}

Second, to have at least $m$ number of tuples for every group with probability $1 - \delta$, \verdict uses the properties of Bernoulli sampling. Let $p_c$ be a sampling probability for the tuples with $c$ in $\mathcal{C}$. Below, we first derive the minimum number of the tuples to be sampled when a certain $p_c$ is used. Our derived expression holds with probability at least $1 - \delta$.
Then, we can compute the value of $p_c$ that ensures the number of the samples in every group to be larger than $m$.

Let $X$ denote the number of the sampled tuples for the group $c$. Since every tuple is sampled independently with probability $p_c$ and there are $|T_{c}|$ such tuples, $X$ follows a Binomial distribution $B(|T_{c}|,\, p_c)$. We want $p_c$ large enough to satisfy $\Pr \left( X \ge m \right) \ge 1 - \delta$. With a standard approximation of $B(|T_{c}|,\, p_c)$ as a normal distribution $N(|T_{c}| \cdot p_c, \;|T_{c}| \cdot p_c \cdot (1 - p_c) )$, we have
\begin{align*}
\int_{m}^{\infty} \frac{1}{\sqrt{2 \pi}}
\exp \left(
  - \frac{(x - |T_{c}| \cdot p_c)^2}{2 |T_{c}| \cdot p_c \cdot (1 - p_c)}
\right)
\ge 1 - \delta
\end{align*}
Then,
\begin{align*}
g(p_c) &= \sqrt{2 |T_c| \cdot p_c (1 - p_c)} \, \erfc^{-1} \left( 2 (1 - \delta) \right) + |T_c| \, p_c
\ge m \\
p_c &\ge g^{-1} (|T_c|, m)
\end{align*}
where $\erfc^{-1}$ is an inverse of the (standard) complementary error function. We numerically found $p_c$ such that $p_c \ge g^{-1} (|T_c|, m)$.

\ph{2. Constructing the Sample}
Let $G$ be a table with schema (\texttt{C}, \texttt{size}) = ($\mathcal{C}$, $|T_c|$). Here, we assume $\mathcal{C}$ contains one column, e.g., \{\texttt{city}\}, to simplify our description. Generalizing the description to the column set with more than one column is straightforward.

When \verdict computes the second candidate sampling probabilities in SQL, it approximates $g^{-1}(|T_c|, m)$ using the values pre-computed for several $m$ and $\delta$. Recall that $m$ is the minimum number of tuples we want to sample for every group. For example, suppose $m = 10$. \verdict first precomputes $g^{-1}(|T_c|, m)$ for $|T_c| = 10, 20, 30, \ldots$. Then, when we need $g^{-1}(25, m)$, \verdict finds the $|T_c|$ values smaller than 25, i.e., 10 and 20. Then, \verdict chooses the largest one among them, i.e., 20. Finally, $g^{-1}(25, m)$ is approximated by $g^{-1}(20, m)$. Since $g^{-1}(|T_c|, m)$ is a strictly decreasing function of $|T_c|$, this approach produces conservative sampling probabilities. The rewritten SQL statement is in the following form:
\begin{lstlisting}[
    basicstyle=\scriptsize\ttfamily,
    xleftmargin=10pt
]
select (column list of T)
from T inner join
     (select C,
             N_T / C_count * tau / size as prob1.
             (case when size > 1000 then ginv(1000)
                   when size > 900  then ginv(900),
                   ...
                   when size > 20   then ginv(20),
                   when size > 10   then ginv(10),
                   else 1.0) as prob2
     from G) as G1 on T.C = G1.C
where rand() < (case when prob1 > prob2 then prob1 else prob2);
\end{lstlisting}
where \texttt{ginv($\cdot$)} stands for the value of $g^{-1}(\cdot, m)$. The \texttt{case} expression in the \texttt{where} clause returns a larger value between the two candidate sampling probabilities, i.e., \texttt{prob1} and \texttt{prob1}. Note that, although \texttt{rand() < prob1 OR rand() < prob2} is simpler, it is an incorrect expression since, in that expression, the \texttt{rand()} functions are evaluated twice and produce two independent random numbers. The situation is same even when \texttt{rand()} is stated in a derived table and its aliased column name is propagated up to the point at which the selection predicate is evaluated.

The actual implementation of \verdict is slightly more involved due to more \texttt{case} expressions inserted for handling the tuples with \texttt{null} values in $\mathcal{C}$. \verdict also computes the actual ratio between the number of the sampled tuples for each group to the original number of the tuples in the respective group; and use those ratios as the values in the sampling probability vector $\vec{p}$. This approach increases the accuracy of AQP answers.
}


\section{Variational Subsampling: Principle}
\label{sec:subsampling}


In this section, we describe \verdict's novel error estimation technique.  
Previous AQP engines,
	especially those that support general analytical queries,
		have relied on bootstrap~\cite{mozafari_sigmod2014_abm},
			which belongs to a family of error estimation techniques called \emph{resampling} \cite{bickel2012resampling,kleiner2012big}.
Resampling techniques, despite various optimizations~\cite{easy_bound_bootstrap,mozafari_sigmod2014_diagnosis}, 
	are still too expensive to be implemented at a middleware layer. 
Therefore,
 we propose the use of a different class of error estimation techniques, called \emph{subsampling},
	for the first time
    in an AQP context. 
Although subsampling is, in general, much more efficient than resampling, 
	direct application of  subsampling theory can be still quite daunting. 

In the remainder of this section,  we  first provide a general overview of subsampling (\cref{sec:subsampling:basics}), and explain why 
	its traditional variant is too expensive.
We then propose a new variant, called \emph{variational subsampling},  which dramatically reduces the cost of 
	traditional subsampling without  compromising its statistical correctness (\Cref{sec:subsampling:basic}). 
Later, in \cref{sec:subsampling:complex}, we generalize this idea to more complex queries, such as nested queries and joins.

\subsection{Subsampling Basics}
\label{sec:subsampling:basics}

Before presenting the basics of subsampling theory,
	we first discuss bootstrap.
Bootstrap is the state-of-the-art error estimation mechanism used by previous AQP engines,
especially those that support general analytical queries~\cite{easy_bound_bootstrap,mozafari_sigmod2014_abm,canty2006diagnostics,hall1988symmetricbootstrap,ranalli2013bootstrap,kleiner2013diagnostic}.
We then discuss why subsampling is more amenable to efficient execution than bootstrap.


\ph{Bootstrap}
Let $g(\cdot)$ be an aggregate function (e.g., mean, sum), which we wish
to compute on $N$ real values $x_1, \ldots, x_N$ (e.g., values of a particular column), 
 i.e., $g(x_1, \ldots, x_N)$. 
 Let a simple random sample of these $N$ values be $X_1, \ldots, X_n$, and $\gs(\cdot)$ be an estimator of $g(\cdot)$.\footnote{For example,
   $\gs(\cdot)$=$g(\cdot)$ 
   	when $g$ is \texttt{avg},  but $\gs(\cdot)$=$\frac{N}{n}g(\cdot)$ when $g$ is \texttt{sum}.}
 That is, we can estimate $g(x_1, \ldots, x_N)$ using $\gs_0 = \gs(X_1, \ldots, X_n)$. 
In an AQP context, we also need to measure the quality (i.e., expected error) of the estimate $\gs(X_1, \ldots, X_n)$.

To measure the quality of the estimate, bootstrap recomputes the aggregate on many resamples, 
	where each resample is a simple random sample (with replacement) of the original sample. 
In bootstrap, the size of a resample is the same as the sample itself, i.e., some of the elements 
$X_1, \ldots, X_n$ might be missing and some might be repeated, but the total number remains as $n$.
 Let $\gs_j$ be the value of the estimator computed on the $j$-th resample, and  $b$ the number of resamples ($b$ is usually a large number, e.g., 
 	100 or 1000).
Bootstrap uses $\gs_1, \ldots, \gs_b$ to construct an empirical distribution of the sample statistics,
	which can then be used to compute a confidence interval. 
Let $\gs_0$ be the estimator's value on the original sample itself, and
 $t_\alpha$ be the $\alpha$-quantile of $\gs_0 - \gs_j$.
 Then, the $1 - \alpha$ confidence interval can be computed as:
\[
\left[ \gs_0 - t_{1 - \alpha/2}\quad,\quad
\gs_0 - t_{\alpha/2} \right]
\]

Due to its generality, 
	bootstrap has been used in many different domains~\cite{mozafari_pvldb2014}.
Although there is an I/O-efficient variant of bootstrap, 
  called \emph{consolidated bootstrap}~\cite{mozafari_sigmod2014_diagnosis},
 its computational overhead remains high, due to the repetitive computation of the aggregate,
	which has a  time complexity of $O(n\cdot b)$. 
Another variant, called \emph{analytical bootstrap}~\cite{mozafari_sigmod2014_abm}, reduces the computational cost but requires modifying the relational operators inside the database (thus, inapplicable to \verdict, which is a middleware).

\ph{Subsampling}
Subsampling follows a procedure similar to bootstrap, 
but with two key differences:
(1)  instead of 
resamples, it uses \emph{subsamples} which are much smaller,
and (2) instead of drawing tuples from the original sample with replacement, 
	subsampling draws tuples without replacement.
In other words, a subsample is also a simple random sample 
	of the original sample, but without replacement,
		and of size $n_s$  where $n_s \ll n$. 
In general, $n_s$ must be chosen such that it satisfies the following two conditions~\cite{politis1994large}: 
	(1) $n_s \rightarrow$ $\infty$ as $n \rightarrow \infty$,
	and (2) $n_s / n \rightarrow 0$ as $n \rightarrow \infty$. 
    Once the subsamples are constructed, 
    	the time complexity of
        the aggregation
          is only $O(n)$; however, constructing the subsamples can itself take $O(b \cdot n)$. We discuss this in more detail shortly. 


Computing the $1 - \alpha$ confidence interval is similar to bootstrap, but requires a scaling:
\[
\left[ \gs_0 - t_{1 - \alpha/2} \cdot \sqrt{n_s / n}\quad,\quad
\gs_0 - t_{\alpha/2}  \cdot \sqrt{n_s / n} \right]
\]
\rev{C3}{In theory, the difference between the empirical confidence interval and the true interval is $O(b^{-1/2} + b/n)$ \cite{politis1994large}.}

While more efficient than bootstrap (since $n_s \ll n$), 
	performing subsampling as a middleware can still be quite expensive.
We illustrate this inefficiency by exploring a  few possible implementations 
	of subsampling in SQL, which is what a middleware would have to do.
(We empirically compare various error estimation techniques in \cref{sec:subsampling:efficiency}.)

\ph{Implementing Subsampling in SQL}
Suppose we need to compute \texttt{sum(price)} of the \texttt{orders} table grouped by \texttt{city}. 
Also, let \texttt{orders\_sample} be a sample table of the \texttt{orders} table. 
One can implement subsampling both with and without User Defined Aggregates (UDAs). 
As a toy example, suppose $n$=1M, $n_s$ = 10K, $b = 100$.

To implement traditional subsampling without UDAs, 
	one needs to first construct a temporary table, say \texttt{orders\_subsamples} which, in addition to the original columns,
		must also have an additional column, say \texttt{sid},  indicating the subsample that each tuple belongs to.
Here, the \texttt{sid} column would contain the subsample id (integers between 1 and $b$).
Given that some tuples might belong to multiple subsamples, the same tuple may appear multiple times, but each time with a different \texttt{sid}.\footnote{Here, we could just keep the aggregation column   instead of the entire tuple.} 
 However, there should be exactly $n_s$ tuples with the same \texttt{sid}. 
 Given such a table, one can use the following query to compute the aggregate on $b$=100 different subsamples (scaling factors omitted for simplicity):
\begin{lstlisting}[
    basicstyle=\footnotesize\ttfamily,
    xleftmargin=5pt,
    caption={Performing traditional subsampling without UDAs.},captionpos=b,
    label={fig:naive_subsampling}
]
select city,
  sum(price * (case when sid = 1 then 1 else 0)),
       ...
  sum(price * (case when sid = b then 1 else 0))
from orders_subsamples
group by city;
\end{lstlisting}
This query costs $O(b \cdot n_s)$,
	but constructing the \texttt{orders\_subsamples} table itself costs $O(b \cdot n)$.\footnote{Note that these 
		subsamples should not be precomputed offline and reused for every query, due to the risk of
     \emph{consistently incorrect estimates}~\cite{mayo1981defense}.}

When the underlying database supports UDAs, one can avoid the need for constructing the \texttt{orders\_subsamples} table,
	and produce the subsamples and their aggregates in a single scan. 
Let \texttt{subsum(price)} be a UDA that, while making a single pass on the original sample, maintains a random subset of exactly $n_s$ tuples using reservoir sampling, and at the end of the scan returns  the sum of the \texttt{price} values for the selected tuples.
Then, one can use the following query, whereby 100 instances of
	the UDA will each return the aggregate value on a separate subsample:


\begin{lstlisting}[
    basicstyle=\footnotesize\ttfamily,
    xleftmargin=5pt,
    caption={Performing traditional subsampling using UDAs.},captionpos=b,
    label={fig:naive_subsampling2}
]
select city,
       subsum(price) as subsample_agg1,
       ...
       subsum(price) as subsample_agg100
from orders_sample
group by city;
\end{lstlisting}
Assuming an ideal case, where the underlying database uses a shared scan among all the UDAs,
	the time complexity is still $O(b \cdot n)$,
		i.e., $b$ UDAs each reading $n$ tuples.


Next, we propose a novel variant of subsampling, which we call \emph{variational subsampling}.
We show that our variant has a time complexity of $O(n)$,  
and is hence much more efficient than traditional subsampling.



\subsection{Variational Subsampling}
\label{sec:subsampling:basic}

In this section, we introduce our new subsampling technique, 	called \emph{variational subsampling},
		which relaxes some of the requirements of traditional subsampling. 
We show that our proposal, while significantly more efficient, still retains the statistical correctness of  traditional subsampling.
In the following subsections, we will generalize our idea to more complex queries.


\ph{Core Idea}
In traditional subsampling, the same tuple \emph{must} be able to belong to multiple subsamples,
	and each subsample \emph{must} be exactly of size $n_s$. 
Enforcing these restrictions is a major source of computational inefficiency.
Our proposed technique relaxes these restrictions, 
	by (1) allowing each tuple to belong to, at most, one subsample,
		and (2) allowing the sizes of different subsamples to differ.
Surprisingly, 
	our analysis reveals that the asymptotic properties of subsampling continue to hold 
		despite lifting these restrictions.
The only caveat is that one must scale the aggregates accordingly (\cref{thm:exclusive}).
However, these two relaxations 
	make a
	 critical difference in terms of computational efficiency: for each tuple, we now only 
	 	need to generate a single random number to determine which subsample it belongs to (if any),
			and then perform the aggregation only once per tuple,
				instead of repeating this process $b$ times.
		\ignore{\footnote{Since \verdict communicates with the underlying database only in SQL, the exact implementation may differ depending on database systems. This single scan analysis is based on the case in which the underlying database implements \texttt{group-by} operations efficiently. The exact performance for different systems are reported in \cref{sec:experiments}.}}

To state this process more formally, we first need to define a \emph{variational table}.
We then explain how to populate this table efficiently.

\begin{definition}
\textbf{(Variational Table)}
Let $b$ be the desired number of subsamples. 
A variational table is a sample table augmented with an extra column 
	that is populated by  random integers between 0 and $b$ (inclusive), 
	 generated independently according to the following weights: $(n - b \cdot n_s,\, n_s,\, n_s,\, ...,\, n_s)$.
	That is, 0 is chosen with probability
	 $\frac{n - b \cdot n_s}{(n - b \cdot n_s) + b \cdot n_s}=\frac{n - b \cdot n_s}{n}$ and each of the integers 1$, \cdots, b$ are
		chosen with probability $\frac{n_s}{(n - b n_s)+ b\cdot n_s}=\frac{n_s}{n}$.
An integer  between 1 and $b$ indicates the subsample id that the current tuple belongs to, whereas 0 indicates that the tuple does not belong to any subsamples.
\label{def:variational_table}
\end{definition}

Note that the independent sampling is an \emph{embarrassingly parallel} process, but it also means that the subsamples are no longer guaranteed to be of size $n_s$. 
Later, 
we show how, with proper scaling, we can still 
	 achieve the same asymptotic properties as those offered by traditional subsampling. 

Nonetheless, 
a variational table can be populated in a straightforward fashion. As we scan the sample table, 
	we randomly assign a single \texttt{sid} (i.e., subsample id) to each tuple. An \texttt{sid} between 1 and $b$ indicates that the tuple belongs to the subsample represented by that integer, while an \texttt{sid} of 0 indicates that the tuple does not belong to any subsample.
Since each tuple is assigned to one subsample at most, this approach effectively partitions the sample into $b$ subsamples, plus the set of those tuples that 
	are not used in any subsample. 
The tuples belonging to different subsamples can be aggregated separately in SQL using a \texttt{group-by} clause.

 
\Cref{fig:basic_var_subsampling} illustrates how this process can be expressed in a single SQL statement, 
	using a toy example, with $n = 10$M, $n_s = 10$K, and  $b$ = 100:
\lstset{escapeinside={<@}{@>}}
\begin{lstlisting}[
    basicstyle=\footnotesize\ttfamily,
    xleftmargin=5pt,
    caption={Example of creating a variational table.},captionpos=b, % with $n$=10M, $n_s$=10K, and $b$=100.}
    label={fig:basic_var_subsampling}
]
select *, 1+floor(rand() * 100) as sid
from orders_sample
where 1+floor(rand() * 1000) <= 100;
\end{lstlisting}
 This query randomly assigns, on average,  $n_s$ tuples to each of the $b$ non-intersecting subsamples. 
To implement the weighted sampling, it uses the expression \texttt{1+floor(rand() * 1000)}, which returns a random integer between 1 and 1000 (inclusive) with equal probability. Values  outside the range $[1,100]$ are treated as 0, and are discarded accordingly. 
	This is because a tuple should not belong to any of the subsamples, with probability
	$\frac{n - b \cdot n_s}{n}=\frac{10M - 100*10K}{10M}=0.9$. 
Once such tuples are discarded, the remaining tuples have  an integer in the range $[1,b]$, representing their \texttt{sid}.
Note that even if the two instances of the \texttt{rand()} function (in the \texttt{select} and \texttt{where} clauses)
	 return different values for the same tuple, the overall probabilities  remain the same.  

\cref{fig:basic_var_subsampling} generates the variational table with $O(n)$ operations, 
	and can be embedded in another query to perform the aggregation on its output.
Below is an example of how to perform the entire variational subsampling in a single query.
\begin{lstlisting}[
    basicstyle=\footnotesize\ttfamily,
    xleftmargin=5pt,
    caption={Example of variational subsampling.},captionpos=b,
    label={fig:var_subsampling}
]
select city, sum(price), count(*) as ns
from (select *, 1+floor(rand() * 100) as sid
       from orders_sample
       where 1+floor(rand() * 1000) <= 100
     ) as orders_v
group by city, sid;
\end{lstlisting}

Note that here we are also returning the size of each subsample (the \texttt{ns} column).
This is because, unlike traditional subsampling, 
	our subsamples might vary in size;
as we discuss
in \cref{thm:exclusive},
 variational subsampling uses these sizes to correct its distribution of the sample estimate. 

Nonetheless, it is easy to see that \cref{fig:var_subsampling} is considerably more efficient than traditional subsampling.
\cref{fig:var_subsampling} performs two aggregates per each of the $b \cdot n_s$ tuples in the \texttt{orders\_v}; thus, the aggregation cost is $O(b\cdot n_s)$. Since the cost of the inner query (building the \texttt{orders\_v})
	is $O(n)$ and $b\cdot n_s \ll n$, the overall time complexity of \cref{fig:var_subsampling} is only
$O(n + b \cdot n_s)$=$O(n)$. 
Therefore, variational subsampling is at least $O(b)$ times more efficient than traditional subsampling, which costs $O(b \cdot n)$ operations
(see \cref{sec:subsampling:basics}).



\ph{Error Correction} To 
guarantee that the distribution of the
	 subsample aggregates converges to the true distribution of the aggregate on the original sample,
	variational subsampling has to correct 
		for   the varying sizes of its subsamples.
Let $n_{s,i}$ denote the size of the $i$-th subsample.
Below, we formally show that the following empirical distribution converges to the true distribution of a sample estimate:
\begin{equation}
L_n(x) = \frac{1}{b} \sum_{i=1}^b \mathds{1} \left( \sqrt{n_{s,i}} ( \gs_i - \gs_0 ) \le x \right)
\label{eq:variationalformula}
\end{equation}
where $x$ is the deviation of a subsample aggregate from the sample aggregate, 
and $\mathds{1}()$ returns 1 if its argument is true; and 0 otherwise.
(Recall that 
$\gs_0$ and $\gs_j$ are the values of the estimator computed on the original sample and the $i$-th resample, respectively.)

\begin{theorem}
Let $J_n(x)$ denote the (non-degenerate) true distribution (cumulative distribution function) of the estimate based on a sample of size $n$.
Then, for any $n_s$ such that $n_s \rightarrow \infty$ and $n_s/n \rightarrow 0$ as $n \rightarrow \infty$,
\[
L_n(x) \rightarrow J_n(x)
\]
in distribution as $n \rightarrow \infty$.
\label{thm:exclusive}
\end{theorem}

This theorem implies that, when $n$ is large, variational subsampling can correctly estimate the distribution of a sample estimate.
%
%
%
The proof to this theorem is presented in \cref{sec:proofs}.
In \cref{sec:convergence}, we show that variational subsampling's asymptotic error is minimized when one chooses $n_s = n^{1/2}$.  
This is why \verdict uses $n_s = n^{1/2}$ as default, but  users can choose different values. 
 In \cref{sec:subsampling:error} and \cref{sec:convergence}, we compare the error of
bootstrap, traditional subsampling, and variational subsampling.

\ignore{Now, we prove \cref{thm:exclusive}. 
This proof relies on the fact that each subsample in a variational table can be regarded as a simple random sample of the population.}

 In this section, we used a simple query to illustrate variational subsampling. 
Next, we show how to obtain a variational table for more complex queries.

\section{Variational Subsampling Advanced}
\label{sec:subsampling:complex}

According to \cref{thm:exclusive}, 
	 as long as we can construct a variational table for a query, we can 
		  correctly estimate the distribution of its sample estimate. 
In this section,
	we extend our core idea from \cref{sec:subsampling} to obtain variational tables for joins (\cref{sec:subsampling:joins}) and nested subqueries (\cref{sec:subsampling:aggregate}).

\subsection{Variational Subsampling for Joins}
\label{sec:subsampling:joins}

 Handling joins is a challenging task for all AQP solutions due to two main problems. 
The first problem is joining (uniform) samples leads to significantly fewer tuples in the output~\cite{random-sampling-joins}. 
The second problem is that joining sampled tables 
	leads to inter-tuple dependence in the output~\cite{wander-join}.
To address the first problem, 
	existing AQP solutions\footnote{Others have resorted to online sampling of the joined relations~\cite{wander-join,ripple_join};
	however, as a middleware, \verdict is currently based on offline sampling.}
		use at most one sampled relation per join~\cite{join_synopses},
		or require the join key to 
			be included in the stratified sample~\cite{mozafari_eurosys2013}  
				or a hashed sample~\cite{kandula2016quickr}. 
\verdict uses the same strategies for 	sidestepping the low cardinality of the join,
	and focuses on solving the second problem.
This is because even when the first problem can be solved by the aforementioned solutions,
	 efficient accounting of the inter-tuple correlations is still a challenge.

To address the second problem, previous solutions have either made   strong foreign-key (FK) assumptions on the join key~\cite{join_synopses}, 
	or have used Horvitz-Thompson (HT) estimators~\cite{kandula2016quickr} and resampling techniques \cite{easy_bound_bootstrap}
	to account for inter-tuple correlations.
As a middleware, \verdict cannot enforce FK relationships,
		and expressing HT estimators for correlations~\cite{kandula2016quickr} in SQL will involve expensive self-joins. 
Also, as mentioned earlier, resampling strategies are too costly for a middleware. 
Instead, \verdict extends its variational subsampling to automatically account for inter-tuple correlations,
	in a manner that can easily be expressed in SQL and efficiently executed by the underlying database. 
From a high-level, to use variational subsampling for a join, we need to construct a variational table of the join output. 
In the rest of this section, we explain how to efficiently obtain a variational table of a join.


Suppose a query involves an aggregation over the join of the \texttt{orders} and \texttt{products} tables, 
	and \verdict decides to use their respective sample tables to compute an approximate output.
To estimate the quality of this approximate answer,
	 \verdict uses the variational tables of the source relations, i.e., \texttt{orders} $\bowtie$ \texttt{products}.

A basic
 approach to constructing a variational table of the join is as follows. 
Given the variational tables of the two tables, i.e., \texttt{orders\_v} and \texttt{products\_v},
	join each subsample of the first table with its corresponding subsample from the other table
	 to construct a new subsample, which we call a \emph{joined subsample}.
	  Repeat this process $b$
	  	 times to construct a variational table of the join query.
The following theorem guarantees the correctness of this approach (see \cref{sec:proofs} for proof).


\begin{theorem}
 Let $g(T, S)$ be an aggregate function involving two tables $T$ and $S$,
and $\gs(T_s, S_s)$ be an estimator of $g(T, S)$, where $T_s$ and $S_s$ are respective samples of $T$ and $S$.
Furthermore,
 let $T_{s,i}$ and $S_{s,i}$ be the $i$-th subsamples of $T_s$ and $S_s$, respectively.
Lastly, let $n_{s,i}$ denote the size of the join of $T_{s,i}$ and $S_{s,i}$.
If $|T_s|/|T_{s,i}| = |S_s| / |S_{s,i}|$, then
$
L(x) = \frac{1}{b} \sum_{i=1}^b \mathds{1} \left( \sqrt{n_{s,i}} \, (\gs(T_{s,i}, S_{s,i}) - \gs(T_s, S_s)) \le x \right)
$
converges to the true distribution of $\gs(T_s, S_s)$ as $n \rightarrow \infty$.
\label{thm:join_subsample}
\end{theorem}

 In other words, the   above theorem states  that we can estimate the distribution 
of our sample-based join approximation, $\gs(T_s, S_s)$, by recomputing the join on 
  respective 
subsamples of $T_s$ and $S_s$, namely 
$T_{s_i}$ and $S_{s_i}$.
 %
However, implementing this approach in SQL would entail a \texttt{union} of multiple join expressions,
	resulting in an extremely inefficient query plan.

In \verdict, we take a significantly more efficient approach, based on a key observation.
We formally show that,
 instead of
 repeatedly
	joining multiple subsamples,
	it suffices to simply join the two variational tables only once, 
		followed by reassigning the
       their \texttt{sid} values
		using a special
 		function (formally introduced in \cref{eq:var_join}).
We show that this approach requires only a single join and a single projection; thus,
it can easily (and efficiently) be implemented in SQL.
To prove the correctness of this approach, we first need to 
	explain the basic approach more formally.

\ph{Basic Approach} To produce a single joined subsample of \texttt{orders} $\bowtie$ \texttt{products}, 
we need to join $\sqrt{b}$ subsamples of \texttt{orders\_v} with $\sqrt{b}$ subsamples of \texttt{products\_v} to produce a joined subsample, where $b$ is the number of the subsamples in each table.

Before a formal presentation, we first use a toy example. 
Suppose \texttt{orders\_v} and \texttt{products\_v} each contain $b$=100 subsamples of size 10K.
Suppose their join, namely \texttt{orders\_v} $\bowtie$ \texttt{products\_v}, has 1M tuples. 
Observe that the probability of two randomly chosen tuples from \texttt{orders\_v} and \texttt{products\_v} satisfying the join condition is 
$\frac{1M \cdot 1M}{1M} = \frac{1}{1M}$.
 We can calculate the number of subsamples needed from
  each of \texttt{orders\_v} and \texttt{products\_v}
	 to yield a joined subsample of size 1M/100 = 10K. 
	Let this number be $x$. Since the join probability is 1/1M, $x$ must satisfy
$
\frac{(x \times 10K) \cdot (x \times 10K)}{1M} = 10K
$. 
This means $x$=10. 
 In this example, we see that $\sqrt{b} = \sqrt{100}$=10=x.

To formally express this process using  relational algebra, let $T_v$ and $S_v$
denote the variational tables of
 two original tables
  $T$ and $S$, respectively.
Further, denote the $i$-th and $j$-th subsamples in those variational tables by $T_{v,i}$ and $S_{v,j}$.
Also, let $\mathcal{I}$ and $\mathcal{J}$  be index sets with integers from 1 to $b$, i.e., 
$\mathcal{I}$=$\mathcal{J}$=$\{1, 2, \ldots, b \}$.
 Then,
\begin{align}
(T \bowtie S)_{v,k}
\; = \; 
\cup_{i \in \mathcal{I}_k} T_{v,i} \; \bowtie \; \cup_{j \in \mathcal{J}_k } S_{v,j} 
\label{eq:join_var1}
\end{align}
where $\mathcal{I}_k$ is a subset of $\mathcal{I}$, $\mathcal{J}_k$ is a subset of $\mathcal{J}$, and each of $\mathcal{I}_k$ and $\mathcal{J}_k$ includes $\sqrt{b}$ elements.
Since $(T \bowtie S)_{v} = \cup_{k=1, \ldots, b}\, (T \bowtie S)_{v,k}$,
the join operation on the right-hand side
  of \cref{eq:join_var1} must be repeated $b$ times.

\begin{figure}[t]

\begin{tikzpicture}

\node[
  anchor=north west,
  draw=mPurple,
  minimum width=15mm,
  minimum height=20mm,
  ultra thick,
] (T) at (0,0) {};

\node [
  fill=white,
  draw=mPurple,
  minimum width=14.8mm,
  minimum height=8mm,
  fill=mPurple,
  fill opacity=0.1,
] at (T) {};

\node[anchor=south west,mPurple,font=\bf] at ($(T.north west)+(0,0.1)$) {$T_v$};


\node at (T) {$\cup_{i \in \mathcal{I}_k} \, T_{v,i}$};
\node[font=\normalsize] at ($(T)+(0,0.3)$) {$\vdots$};
\node[font=\normalsize] at ($(T)+(0,-0.5)$) {$\vdots$};

\draw [decorate,decoration={brace,amplitude=10pt,mirror},xshift=-4pt,yshift=0pt,mPurple]
($(T.north west)+(-0.2,0)$) -- ($(T.south west)+(-0.2,0)$)
 node [midway,xshift=-0.8cm,align=center,font=\footnotesize] {$\sqrt{b}$\\ subsets};


\node[
  anchor=north west,
  draw=black,
  minimum width=40mm,
  minimum height=20mm,
  ultra thick,
  font=\footnotesize,
] (J) at (2.3,0) {};

\node[
  draw=black,
  minimum width=16mm,
  minimum height=8mm,
  fill=gray,
  fill opacity=0.1,
] (c) at (J) {};
\node at (c) {$(T \bowtie S)_{v,k}$};

\node[
  align=left,
  anchor=south,
  font=\footnotesize,
] (c2) at ($(c.north)+(0,0.1)$) {One of $b$ ($k$-th) joined subsamples};

\node[anchor=south west,black,font=\bf] at ($(J.north west)+(0,0.1)$) {$(T \bowtie S)_v$};

\node[
  anchor=north west,
  draw=darkorange,
  minimum width=40mm,
  minimum height=10mm,
  ultra thick,
] (S) at ($(J.south west)+(0,-0.3)$) {};

\node [
  fill=white,
  draw=darkorange,
  minimum width=16mm,
  minimum height=9.8mm,
  fill=darkorange,
  fill opacity=0.1,
] at (S) {};

\node[anchor=north east,darkorange,font=\bf] at ($(S.north west)+(-0.1,0)$) {$S_v$};


\node[] at (S) {$\cup_{j \in \mathcal{J}_k} \, S_{v,j}$};
\node[font=\Large] at ($(S)+(-1.3,0)$) {$\cdots$};
\node[font=\Large] at ($(S)+(1.3,0)$) {$\cdots$};

\draw [decorate,decoration={brace,amplitude=10pt,mirror},xshift=-4pt,yshift=0pt,darkorange]
($(S.south west)+(0,-0.2)$) -- ($(S.south east)+(0,-0.2)$)
 node [midway,yshift=-0.5cm] {\footnotesize $\sqrt{b}$ subsets};

\draw [dashed] ($(T.west)+(1.5,0.4)$) -- (c.north west);
\draw [dashed] ($(T.west)+(1.5,-0.4)$) -- (c.south west);
\draw [dashed] ($(S.north)+(-0.8,0)$) -- (c.south west);
\draw [dashed] ($(S.north)+(0.8,0)$) -- (c.south east);

\end{tikzpicture}

\vspace{-2mm}

\caption{Joining variational tables to construct a new variational table of a join.
Instead of repeatedly
  joining $\sqrt{b} \times \sqrt{b}$ pairs of subsamples, 
  we simply join  the two variational tables (only once), then reassign  their \texttt{sid} values.}
\label{fig:join_var}
\end{figure}
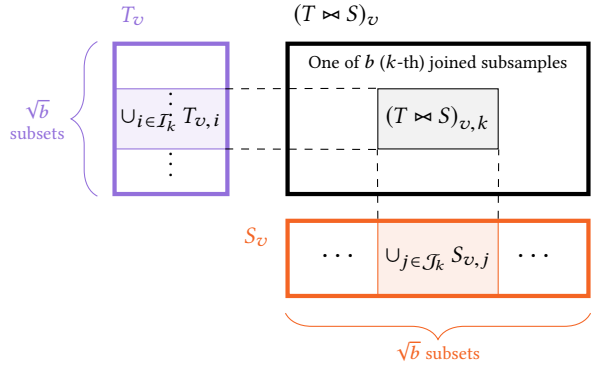

\ph{Efficient Approach}
Our key observation is that the variational table of the join, or equivalently, the union of $(T \bowtie S)_{v,k}$ for $k = 1, \ldots, b$, can be 
converted into a logically equivalent but computationally more efficient expression, if
    the cross product of $\mathcal{I}_k$ and $\mathcal{J}_k$ partitions $\mathcal{I} \times \mathcal{J}$; that is,
\[
\mathcal{I} \times \mathcal{J}
\; = \; 
\bigcup_k \; (\mathcal{I} \times \mathcal{J})_k
\; = \; 
\bigcup_k \; \mathcal{I}_k \times \mathcal{J}_k
\]
We state our result formally in the following theorem (proof deferred to \cref{sec:proofs}).
\begin{theorem}
If there exists
\begin{align}
h(i,j) = k \quad \text{for } (i,j) \in (\mathcal{I} \times \mathcal{J})_k = \mathcal{I}_k \times \mathcal{J}_k
\label{eq:var_join}
\end{align}
\begin{align}
\text{then} \qquad \quad
(T \bowtie S)_{v} = \Pi_{*,\, h(i,j) \; \texttt{as}\;  \texttt{sid}} \; \left( T_{v} \; \bowtie \; S_{v} \right) \qquad \quad
\label{eq:var_join_rel}
\end{align}
\label{thm:join}
\end{theorem}

In this theorem, the projection does not remove duplicates, and the $*$ subscript in the projection means that we preserve all of the existing columns. 
The final projection with ``$h(i,j)$ as \texttt{sid}'' effectively
  identifies the subsamples in $(T \bowtie S)_{v}$, namely, the variational table for $T \bowtie S$.
 Note that, given such an $h(i,j)$ function, the expression in \cref{eq:var_join_rel} can easily be expressed in SQL.

We give an example of the function $h(i,j)$ in \cref{eq:var_join}:
\[
h(i,j) = \floor*{\frac{i-1}{\sqrt{b}}} \cdot \sqrt{b} + \floor*{\frac{j-1}{\sqrt{b}}} + 1
\qquad
i, j = 1, \ldots, b
\]
where $\floor{\cdot}$ returns the floor of its argument. Note that this $h(i,j)$ function is similar to how two-dimensional arrays are indexed sequentially in most programming languages (e.g., C).




\Cref{fig:join_var} visually explains our approach. Sets of subsamples from $T_v$ and $S_v$ are joined to produce the $k$-th joined subsample. Each set contains $\sqrt{b}$ subsamples, and there are $\sqrt{b} \cdot \sqrt{b} = b$ combinations.
 Thus, joining every pair of sets (of subsamples)
   produces $b$ joined subsamples in total. 
  Since the hash function $h(i,j)$ can
 identify $k$ given $i$ and $j$, we can simply join all tuples first, and then assign new \texttt{sid} values.


\subsection{Variational Subsampling for Nested Queries}
\label{sec:subsampling:aggregate}

To illustrate how \verdict obtains a variational table for nested queries, 
 consider the following query as an example:
\begin{lstlisting}[
    basicstyle=\footnotesize\ttfamily,
    xleftmargin=5pt,
    caption={An aggregate query in the from clause.},captionpos=b,
    label={fig:aggregate_source}
]
select avg(sales) as avg_sales
from (select city, sum(price) as sales
      from orders
      group by city) as t;
\end{lstlisting}
For variational subsampling, we need a variational table of \texttt{t}, which we denote by \texttt{t\_v}.

Note that \texttt{t\_v} should be a union of $b$ aggregate statements, where each   aggregate statement  is computed on a subsample. Let a variational table of \texttt{orders} be \texttt{orders\_v} (which includes an \texttt{sid} column to indicate the subsample that each tuple belongs to). Then, a basic approach to obtaining \texttt{t\_v} is
\begin{lstlisting}[
    basicstyle=\footnotesize\ttfamily,
    xleftmargin=5pt,
    caption={A basic approach to obtaining a variational table of \texttt{t}.},captionpos=b,
    label={fig:aggregate_var1}
]
select city, sum(price) as sales, avg(1) as sid
from orders_v
where sid = 1
group by city
union
...
union
select city, sum(price) as sales, avg(b) as sid
from orders_v
where sid = b
group by city;
\end{lstlisting}

However, by exploiting the property that the subsamples in \texttt{orders\_v} are disjoint, 
	we can perform the above operations more efficiently. Formally, let $T_v$ be a variational table. Then,
\begin{align}
\begin{split}
\bigcup_k \; {}_{G}\, {\mathcal{G}}_g \, (T_{v,k}) 
&= \bigcup_k \; {}_{G}\, \mathcal{G}_g \, \left( \sigma_{\texttt{sid} = k} (T_v) \right) \\
&= {}_{G, \texttt{sid}}\, \mathcal{G}_g \, (T_v)
\end{split}
\label{eq:nested}
\end{align}
where ${}_{G}\, {\mathcal{G}}_g$  
 is an aggregate operator with a set of grouping attributes $G$ and an aggregate function $g$. 
Also, $T_{v,k}$ is the $k$-th subsample of $T_v$, as defined in \cref{sec:subsampling:joins}.

\Cref{eq:nested}  indicates that \cref{fig:aggregate_var1}, i.e., the variational table of \texttt{t}, can be alternatively expressed  using the
 variational table of \texttt{orders}, i.e., \texttt{orders\_v}: 

\begin{lstlisting}[
    basicstyle=\footnotesize\ttfamily,
    xleftmargin=5pt,
    caption={A variational table of an aggregate statement.},captionpos=b,
    label={fig:aggregate_subsample}
]
select city, sum(price) as sales, sid
from orders_v
group by city, sid;
\end{lstlisting}

Finally, \cref{fig:aggregate_subsample} can be used in place of \texttt{t} in \cref{fig:aggregate_source} for estimating the quality of a sample estimate. 
Note that \cref{fig:aggregate_subsample} requires $O(b)$ fewer scans (of \texttt{orders\_v}) than \cref{fig:aggregate_var1}.

\ignore{
\barzan{the rest of this subsection is awful. 1) makes no sense, 2) is sloppy, 3) sounds too trivial all at the same time}
Observe that \texttt{t} can be regarded as a relation with the schema (\texttt{city}, \texttt{sales}). Thus, a variational table of $\texttt{t}$ is a union of $b$ number of such relations, each of which is computed on a subsample.
Thus, given a variational table \texttt{orders\_var\_table} of \texttt{orders}, we can express the variational table of \texttt{t} as follows:

Observe that there are $b$ subsamples (or $b$ \texttt{sid}) for each value of \texttt{city}. Since the outer query is aggregating the sales of all cities, we can get $b$ number of those aggregations. Those $b$ number of aggregations, then, are used for estimating the quality (or equivalently, the distribution) of the outer aggregate computed on a sample.
}

\ignore{
Recall that in our earlier example (i.e., \cref{fig:var_subsampling}), to estimate the quality of an aggregate function in the outer-most query (i.e., \texttt{sum(price)}), we needed a variational table its source table (i.e., \texttt{orders}). The same is true here. To estimate the quality of an aggregate function in the outer-most query (i.e., \texttt{avg(sales)}), we need a variational table of its source table (i.e., $\texttt{t}$). Once the variational table of \texttt{t} is obtained, the rest of the procedure is identical.

We can convert the source table (i.e., \texttt{t}) to a variational table for some important cases (\cref{sec:nested_var}). 
This conversion has a few requirements. First, the output of the converted table must have the \texttt{sid} column, so that it satisfies the definition of variational table (\cref{def:variational_table}). Also, the tuples that belong to different subsamples must not be aggregated together. The following query produces a variational table of \texttt{t} that achieves those two requirements. Notably, this query produces it by performing the same kind of SQL operations on the variational table (i.e., \texttt{orders\_variational\_table}) of its source (i.e., \texttt{orders}). 
The correctness of this approach is shown in \cref{thm:correctness}.

This statement, which expresses the variational table of \texttt{t}, can be used for computing our original nested query as follows.

\begin{lstlisting}[
    basicstyle=\footnotesize\ttfamily,
    xleftmargin=0pt,
    caption={Variational subsampling for an aggregate statement},captionpos=b,
    label={fig:agg_subsampling}
]
select avg(sales) as avg_sales
from ( Query 6 ) as <@\textcolor{red}{t\_variational\_table}@>
group by sid;
\end{lstlisting}
}






\begin{figure*}[t]
\centering
\begin{tikzpicture}
    \begin{axis}[
        speedups,clip=false,ymax=100,
        ytick={0, 50, 100, 150, 200},
        yticklabels={0$\times$,50$\times$,100$\times$,150$\times$,200$\times$},
        ]
        \addplot[fill=gray,draw=gray,point meta = explicit symbolic]
        table[meta=z] {
        x y z
        1	54.66935484	54.67$\times$
        2	1	1.00$\times$
        3	42.14962121	42.15$\times$
        4	34.73441109	34.73$\times$
        5	22.45493562	22.45$\times$
        6	4.845190721	4.85$\times$
        7	17.99972169	18$\times$
        8	1	1.00$\times$
        9	35.22442019	35.22$\times$
        10	10.92731947	10.93$\times$
        11	2.854125399	2.85$\times$
        12	19.05582923	19.06$\times$
        13	1	1.00$\times$
        14	1.968916082	1.97$\times$
        15	107.0647241	107$\times$
        16	2.192143493	2.19$\times$
        17	20.75210292	20.75$\times$
        18	1.056508208	1.06$\times$
        19	8.412698413	8.41$\times$
        20	15.76419214	15.76$\times$
        21	19.34782609	19.35$\times$
        22	18.32618026	18.33$\times$
        23	20.72072072	20.72$\times$
        24	18.04020101	18.04$\times$
        25	3.691860465	3.69$\times$
        26	34.20792079	34.21$\times$
        27	32.4824356	32.48$\times$
        28	27.88104089	27.88$\times$
        29	36.28726287	36.29$\times$
        30	38.21428571	38.21$\times$
        31	16.97142857	16.97$\times$
        32	67.28043776	67.28$\times$
        33	52.27180966	52.27$\times$
        };

    \end{axis}
\end{tikzpicture}
\vspace{-4mm}

\caption{\rev{C6}{\verdict's speedups for Redshift.
 Associated errors are in \cref{fig:exp:error1}.
(Spark and Impala deferred to \cref{fig:exp:speedups2}.)}
 }
\label{fig:exp:speedups}
\end{figure*}
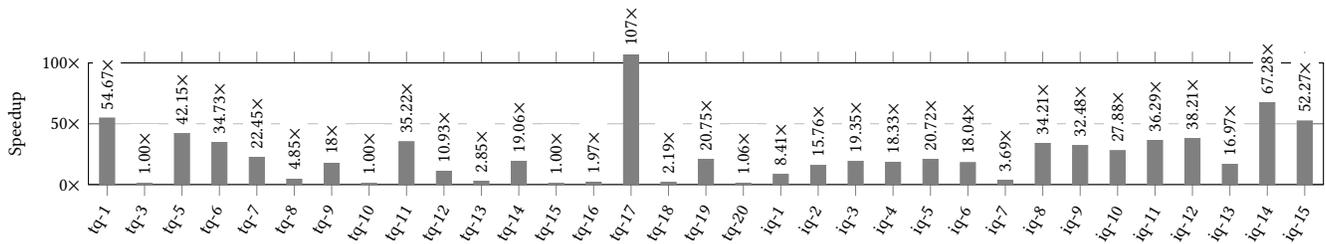



\section{Experiments}
\label{sec:experiments}


In this section, we empirically evaluate \verdict. Our experiments aim to demonstrate \verdict's platform-independence,
 efficiency, and statistical correctness.
In summary, our experiments show the following:

\begin{enumerate}[1.,nolistsep,noitemsep,leftmargin=10pt]
\item Thanks to its UAQP,
	\verdict delivered  an average of 18.45$\times$---and up to 171$\times$---speedup
		(for Impala, Spark SQL, and Redshift),
		 	\rev{C9}{and with less than 2.6\% relative error.} (\cref{sec:speedups}, \cref{sec:errors})

\item \verdict's performance was comparable to (and sometimes even faster than) a tightly-integrated,
	commercial AQP engine, i.e, \snappydata.\footnote{In our experiments, we used \snappydata's community edition version 0.8 (\snappydata's more recent versions are likely to perform better than this version).} (\cref{sec:aqp_comparison})

\item Variational subsampling was 348$\times$ faster than traditional subsampling and 239$\times$ faster than consolidated bootstrap~\cite{mozafari_sigmod2014_diagnosis} expressed in SQL. (\cref{sec:subsampling:efficiency})

\item Variational subsampling yielded statistically correct estimates. (\cref{sec:subsampling:error})


\end{enumerate}
For interested readers, \cref{sec:experiments:preparation} offers additional experiments on \verdict's offline sample preparation overhead.
%
%

\subsection{Setup}
\label{sec:experiments:setup}



\ph{SQL Engines and Clusters}
We used Spark 1.6.0  and Impala 2.8.0 included in CDH 5.11.2.
For Spark SQL and Impala experiments,
we used 10 EC2 \texttt{r4.xlarge} instances as workers and another one as a master.
Each instance had  Intel Xeon E5-2686 v4 processors (4 cores), 30.5 GB memory, and 500 GB SSD for HDFS.
 For Redshift experiments,
	we used 20 \texttt{dc1.large} instances as workers and an additional one as a master.
Each instance had a CPU with 2 cores, 15 GB memory, and 160 GB SSD.

\ph{Datasets and Queries} We used three datasets:
\begin{enumerate}[1.,nolistsep,noitemsep,leftmargin=10pt]
\item \instacart~\cite{instacart}: This is a 100$\times$ scaled sales database of an actual online grocery store called Instacart. The size of the dataset was 124 GB before compression.
\item \tpch~\cite{tpch}: This is a 500 GB standard TPC-H dataset.
\item \synthetic: This is a synthetic dataset we generated to fine-control various properties of data (defined in \cref{sec:subsampling:error}).
%
\end{enumerate}

\noindent
\rev{C7}{Spark SQL and Impala loaded and processed
the Parquet-compressed data from an SSD-backed HDFS;
Amazon Redshift automatically stored them in a compressed columnar format.}

\verdict created sample tables for large fact tables: 1\% uniform samples, 1\% universe samples, and up to 80\% budget for stratified samples.
(We used a larger budget for stratified samples since the \tpch dataset included many high-cardinality columns.)

We used 33 queries in total: 18 out of the 22 \tpch queries\footnote{One query (tq-2) had no aggregates, and the other three included an \texttt{EXISTS} condition, which \verdict currently does not  support.
tq-4 and tq-20 are not supported by previous  AQP engines either \cite{mozafari_sigmod2014_abm,mozafari_eurosys2013}.}
  (numbered
as tq-\# where \# is the \tpch query number~\cite{tpch})
plus 15 micro-benchmark queries on the \instacart dataset (numbered as  iq-1, $\ldots$, iq-15).
The micro-benchmark queries consisted of various aggregate functions on up to 4 joined tables.
We used low-cardinality columns (up to 24, randomly chosen) in the grouping attributes of these micro-benchmark queries.


\subsection{\verdict's Speedup  for Various Engines}
\label{sec:speedups}

This section compares the query latencies of Redshift, Spark SQL, and Impala with and without \verdict.
Since \verdict performs AQP, their query latencies with \verdict are expected to be lower.
However, the purpose of this section is to (1) quantify the extent of the speedup that \verdict can deliver as a UAQP running on top of existing platforms, and
	(2) verify \verdict's ability in supporting common forms of OLAP queries.
Moreover, testing \verdict with several different engines sheds light on the characteristics of a SQL engine that are favorable for UAQP.

We ran each of the 33 queries on Redshift with and without \verdict, and measured their query latencies. We repeated the same process for Spark SQL and Impala.
\Cref{fig:exp:speedups} reports \verdict's speedups for Redshift, i.e., the latency of the regular engine divided by \verdict's latency.
For 3 out of the 18 \tpch queries (tq-3, tq-8, and tq-15), \verdict determined that AQP was not feasible due to the high cardinality of
 the grouping attributes; thus, \verdict simply ran the original queries (i.e., no speedup).
 For other queries, \verdict yielded 1.05$\times$--107$\times$ speedups, with an average speedup of 24.0$\times$. \rev{C9}{The associated errors were less than 2.6\% for all queries (per-query errors are reported in \cref{fig:exp:error1}).}
 The average speedups for Spark SQL and Impala were 12.0$\times$ and 18.6$\times$, respectively
 (their detailed results are deferred to \cref{fig:exp:speedups2}, due to space limitation).

\begin{figure}[t]
    \pgfplotsset{scalability/.style={
        width=75mm,
        height=32mm,
        ymax=40,
        ymin=0,
        xmin=0.5,
        xmax=4.5,
        xlabel=Original Data Size,
        ylabel=Speedup,
        ylabel near ticks,
        ylabel style={align=center},
        xticklabel style={yshift=-2mm},
        legend style={
            at={(0,1)},anchor=north west,column sep=2pt,
            draw=black,fill=white,line width=.5pt,
            /tikz/every even column/.append style={column sep=20pt}
        },
        legend cell align={left},
        legend columns=1,
        every axis/.append style={font=\footnotesize},
        ymajorgrids,
        minor grid style=lightgray,
        every node near coord/.append style={font=\footnotesize,fill=white},
        clip=false,
    }}

\begin{tikzpicture}
    \begin{axis}[scalability,
    ymax=100,
    ymin=1,
    xtick={1,2,3,4},
    xticklabels={5GB, 50GB, 200GB, 500GB},
    ytick={1, 10, 100},
    yticklabels={1$\times$, 10$\times$, 100$\times$},
    ymode=log,
    ]
        \addplot[mark=*,mark size=1.5,black!20!mPurple,
        line width=1mm,point meta = explicit symbolic,
        every node near coord/.append style={anchor=north west,fill=white,text=black!80!mPurple,font=\bf\small},]
        table[x=x,y=y,meta=z] {
        x y z
        1 1.183261183 1.2$\times$
        2 1.630591631 1.6$\times$
        3 6.789321789 6.8$\times$
        4 31.21212121 31.2$\times$
        };

        \addplot[mark=*,mark size=2,black!20!darkorange,
        line width=1mm,point meta = explicit symbolic,
        every node near coord/.append style={anchor=south east,fill=white,text=black!80!darkorange,font=\bf\small},
        ]
        table[x=x,y=y,meta=z] {
        x y z
        1 0.9509707141 0.9$\times$
        2 1.121256992 1.1$\times$
        3 7.202698256 7.2$\times$
        4 14.14939125 14.1$\times$
        };



        \addlegendentry{tq-6}
        \addlegendentry{tq-14}

    \end{axis}
\end{tikzpicture}
\caption{\rev{C7}{Speedups for different data sizes using two queries (sample fixed to 5GB; Impala).} 
}
\label{fig:exp:speedup:scale}
\end{figure}
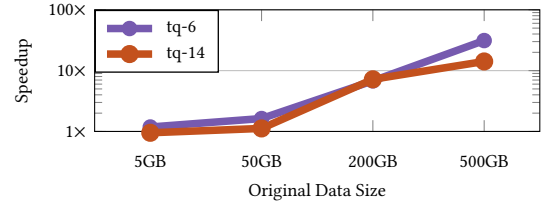

\begin{figure*}[t]
\centering
\pgfplotsset{aqp/.style={
        width=95mm,
        height=32mm,
        ybar=0.8mm,
        bar width=1.2mm,
        ymax=80,
        ymin=0,
        xmin=0.5,
        xmax=3.5,
        ylabel=Runtime,
        xlabel near ticks,
        ylabel near ticks,
        ylabel style={align=center},
        xtick={1,2,...,15},
        xticklabels={tq-1,tq-5,tq-6,tq-7,tq-8,tq-9,tq-11,tq-12,tq-13,tq-14,tq-15,tq-16,tq-17,tq-18,tq-19},
        xticklabel style={rotate=30,xshift=0mm,yshift=1mm},
        ytick={0, 30, ..., 180},
        yticklabels={0 sec,30 sec,60 sec,90 sec,120 sec, 150 sec,180 sec},
        legend style={
            at={(1.0,1.0)},anchor=north east,column sep=2pt,
            draw=black,fill=white,font=\footnotesize,line width=.5pt,
            /tikz/every even column/.append style={column sep=10pt},
        },
        legend columns=2,
        every axis/.append style={font=\footnotesize},
        ymajorgrids,
        yminorgrids,
        minor grid style=lightgray,
        legend cell align={left},
    }}

\begin{tikzpicture}
    \begin{axis}[aqp,clip=false,xmin=0.5,xmax=15.5,ymin=0,ymax=180]
        \addplot[fill=gray,draw=gray]
        table[x=x,y=y] {
        x y
        1 25.5
        2 165.4
        3 11.9
        4 130.4
        5 170.3
        6 74.3
        7 8.8
        8 60.8
        9 7.98
        10 31.3
        11 16.9
        12 20.1
        13 32.0
        14 60.5
        15 39.7
        };

        \addplot[fill=darkorange,draw=darkorange]
        table[x=x,y=y] {
        x y
        1 18.7
        2 43
        3 8.3
        4 35
        5 155
        6 54.7
        7 16
        8 7.4
        9 17.15
        10 27.3
        11 23.7
        12 59.7
        13 32.7
        14 13.0
        15 22.7
        };

        \addlegendentry{\snappydata}
        \addlegendentry{\verdict}
    \end{axis}
\end{tikzpicture}
~
\begin{tikzpicture}
    \begin{axis}[aqp,clip=false,xmin=12.5,xmax=27.5,width=90mm,ymin=0,ymax=20,
    xtick={13,14,...,27},
    xticklabels={iq-1,iq-2,iq-3,iq-4,iq-5,iq-6,iq-7,iq-8,iq-9,iq-10,iq-11,iq-12,iq-13,iq-14,iq-15},
    ylabel=\empty,
    ytick={0,5,...,20},
    yticklabels={0 sec,5 sec,10 sec,15 sec,20 sec},
    legend style={
        at={(0.0,1.0)},anchor=north west,column sep=2pt,
        draw=black,fill=white,font=\footnotesize,line width=.5pt,
        /tikz/every even column/.append style={column sep=10pt},
    },
    ybar=0.8mm,
    bar width=1.2mm,
    ]
        \addplot[fill=gray,draw=gray]
        table[x=x,y=y] {
        x y
        13  1.302
        14  0.456
        15  0.508
        16  1.075
        17  0.711
        18  0.438
        19  0.566
        20  1.059
        21  0.579
        22  1.169
        23  0.606
        24  0.455
        25  0.208
        26  16.269
        27  16.807
        };

        \addplot[fill=darkorange,draw=darkorange]
        table[x=x,y=y] {
        x y
        13  0.89
        14  0.65
        15  0.73
        16  1.48
        17  0.85
        18  0.53
        19  1.11
        20  0.72
        21  1.11
        22  2.47
        23  0.96
        24  0.84
        25  0.30
        26  5.58
        27  4.80
        };

        \addlegendentry{\snappydata}
        \addlegendentry{\verdict}
    \end{axis}
\end{tikzpicture}
\vspace{-6mm}

\caption{\rev{C2}{AQP performance of \verdict vs. \snappydata. \verdict was faster for the queries including joins of samples.}}
\label{fig:exp:aqp}
\end{figure*}
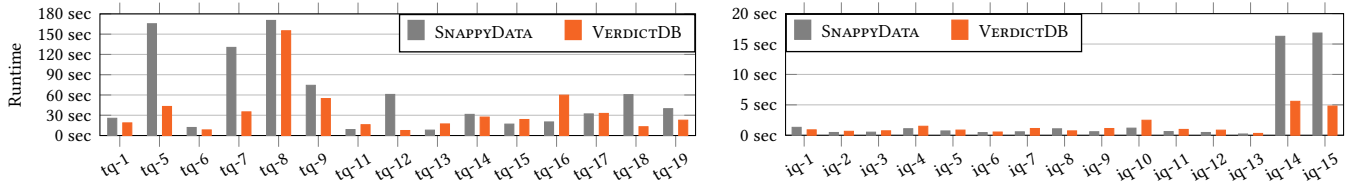

Across these three engines, the speedups were larger when the default overhead of the original engine (e.g., reading the catalog)
was a smaller portion of the overall query processing time.
This is because \verdict (and AQP in general) reduces the data processing time, not
the query preparation time.
This depended on two factors: the default overhead, and the data preparation time.
	\verdict brought a larger speedup when the engine spent less time on catalog access and query planning (e.g.,
		larger speedup for Redshift than Spark SQL).
Likewise, when the engine processed less prepared data,
	\verdict's speedups were more dramatic (e.g., csv file\footnote{When we ran the same set of queries on Impala and Spark with csv files, we observed 56.9$\times$ average speedups.} versus parquet format).

\rev{C8}{Next, we also measured the speedups for different ratios of the sample size to the original data size.
Specifically, we used a fixed sample size of 5 GB,
	while varying the size of the original data from 5 GB to 500 GB. \cref{fig:exp:speedup:scale} depicts the results for two
	queries: tq-6 and tq-14. As expected, when the data size is already small (i.e., 50 GB), there was less room for speedup, i.e., only 1.4$\times$ on average; however, the speedup increased for larger data sizes: 7.00$\times$ for 200 GB and more than 22.6$\times$ for 500 GB.}


\subsection{UAQP versus Tightly Integrated AQP}
\label{sec:aqp_comparison}

This section compares the query latencies of \verdict, as the first example of UAQP, to tightly-integrated AQP systems. \rev{C1}{We first compare \verdict to a tightly-integrated sampling-based AQP engine, \snappydata; then, we compare \verdict to non-sampling-based (adhoc) AQP features natively offered by commercial engines (e.g., HyperLogLog implementation of \texttt{count-distinct}).}

Due to its generality, middleware architecture, and sole reliance on SQL-based computations,
 	\verdict is expected to be slower than tightly-integrated AQP engines that are highly specialized for a particular query engine
	(e.g., \snappydata).
However, our goal here is to understand the extent to which \verdict has traded off raw performance  in exchange
	 for greater generality and deployability.



First, we compared \verdict on Spark SQL against \snappydata. \snappydata is tightly-integrated into Spark SQL.
For these experiments,
we ran the same set of \tpch and \instacart queries.\footnote{We excluded three \tpch queries (i.e., tq-3, tq-10, and tq-20) due to \snappydata's failure in creating the samples stratified on extremely high-cardinality columns.}
\cref{fig:exp:aqp}
reports the per-query latencies.
For most queries, \verdict's performance was comparable to \snappydata. However, there were several queries (i.e., tq-5, tq-7, tq-12, iq-14, iq-15) for which \verdict was significantly faster.
This is because those queries included joins of two samples.
Unlike \verdict, \snappydata does not support the join of two samples (even when the join key is included in a stratified or hashed sample).
In those situations, \snappydata simply used the original table for the second relation, while
	\verdict relied on its hashed samples.

\begin{table}[t]
\small
\begin{subfigure}[b]{\linewidth}
\centering
\begin{tabular}{r r r r}
\toprule
\verdictshort+Impala & Impala & \verdictshort+Redshift & Redshift \\
\midrule
1.1 sec (0.01\%) & 17.1 sec (3.4\%) & 0.5 sec (0.02\%) & 7.7 sec (5.0\%) \\
\bottomrule
\end{tabular}
\vspace{1mm}
\caption{approximate count-distinct runtime and relative error}
\end{subfigure}

\vspace{1mm}

\begin{subfigure}[b]{\linewidth}
\centering
\begin{tabular}{r r r r}
\toprule
\verdictshort+Impala & Impala & \verdictshort+Redshift & Redshift \\
\midrule
1.5 sec (0\%) & 53.2 sec (0\%) & 1.0 sec (0\%) & 106.6 sec (0\%) \\
\bottomrule
\end{tabular}
\vspace{1mm}
\caption{approximate median runtime and relative error}
\end{subfigure}

\vspace{1mm}
\caption{\rev{C1}{Sampling-based AQP vs.~native approximation.}
}
\label{fig:exp:aqp2}
\vspace{-2mm}
\end{table}

 Second, we compared \verdict's sampling-based approximations for \texttt{count-distinct} and median
 against Impala and Redshift's native approximate aggregates (i.e., \texttt{ndv}, \texttt{approx\_median}, \texttt{percentile\_disc}).
\cref{fig:exp:aqp2} summarizes the results.
On average, \verdict's sampling-based results were 43.5$\times$ faster  than the native approximations.
This is because Impala and Redshift's approximate aggregates
rely on sketching techniques that require a full scan over data. As such, their disk I/O cost is higher.


In summary, this experiment confirms that \verdict's much greater generality (i.e., UAQP)
 	comes at only a negligible loss of performance compared to
 	tightly-integrated AQP systems.


\subsection{Variational Subsampling: Efficiency}
\label{sec:subsampling:efficiency}

In this section, we compare the runtime overhead of three resampling-based error estimation methods: consolidated bootstrap, traditional subsampling, and variational subsampling. First, we ran three types of queries (flat, join, and nested) without any error estimation.
We then ran each query with each of these three error estimation methods. By subtracting the query latencies without error estimation,
	we derived the runtime overhead of each error estimation technique.

\cref{fig:exp:subsampling:overhead} reports the query latencies.  Both consolidated bootstrap and traditional subsampling yielded substantial runtime overhead.
Recall that their time complexities are $O(b \cdot n)$.
\rev{C5}{In contrast, variational subsampling added only 0.38--0.87 seconds to the latency of the queries.
The latency overhead comes from sample planning (26 ms on average) and extra \texttt{groupby} and aggregation processes inserted for performing variational subsampling.}
Compared to consolidated bootstrap (which is the state-of-the-art error estimation strategy~\cite{mozafari_sigmod2014_diagnosis,G-OLA}), variational subsampling was 189$\times$, 237$\times$, and 100$\times$ faster, respectively.
Considering the overall query latencies---including the cost of computing the approximate answers and their error bounds---running queries with variational subsampling was 99$\times$, 42$\times$, and 63$\times$ faster than consolidated bootstrap for flat, join, and nested queries,
 respectively.




\begin{figure}[t]
\centering
\begin{tikzpicture}
    \begin{axis}[
        height=32mm,
        ybar=2.5mm,
        width=85mm,
        xtick={1,2,3},
        xticklabels={Flat queries,Join queries,Nested queries},
        xmin=0.5,
        xmax=3.5,
        ymin=0,
        ymax=300,
        ylabel=Query Latencies,
        ytick={0,100,200,300},
        yticklabels={0 sec,100 sec,200 sec,300 sec},
        bar width=3mm,
        legend style={
            at={(-0.05,1.1)},anchor=south west,column sep=2pt,
            draw=black,fill=none,font=\footnotesize,line width=.5pt,
            /tikz/every even column/.append style={column sep=10pt}
        },
        every axis/.append style={font=\footnotesize},
        ymajorgrids,
        yminorgrids,
        minor grid style=lightgray,
        legend columns=2,
        legend cell align={left},
        nodes near coords,
        every node near coord/.append style={font=\scriptsize,fill=white},
    ]

        \addplot[fill=mPurple,draw=mPurple,point meta = explicit symbolic,
        every node near coord/.append style={xshift=0mm}]
        table[x=x,y=y,meta=z] {
        x y z
        1 72 72s
        2 171 171s
        3 88 88s
        };

        \addplot[fill=gray,draw=gray,point meta = explicit symbolic,
        every node near coord/.append style={xshift=0mm}]
        table[x=x,y=y,meta=z] {
        x y z
        1 118 118s
        2 247 247s
        3 124 124s
        };

        \addplot[fill=darkorange,draw=darkorange,point meta = explicit symbolic,
        every node near coord/.append style={xshift=0mm}]
        table[x=x,y=y,meta=z] {
        x y z
        1 0.73 0.73s
        2 4.03 4.03s
        3 1.40 1.40s
        };

        \addplot[fill=black,draw=black,point meta = explicit symbolic,
        every node near coord/.append style={xshift=0mm}
        ]
        table[x=x,y=y,meta=z] {
        x y z
        1 0.35 0.35s
        2 3.33 3.33s
        3 0.53 0.53s
        };

        \addlegendentry{Consolidated Bootstrap}
        \addlegendentry{Traditional Subsampling}
        \addlegendentry{\textbf{Variational Subsampling}}
        \addlegendentry{No error estimation (baseline)}
    \end{axis}
\end{tikzpicture}
\vspace{-2mm}

\caption{Runtime with different error estimation methods.}
\label{fig:exp:subsampling:overhead}
\end{figure}
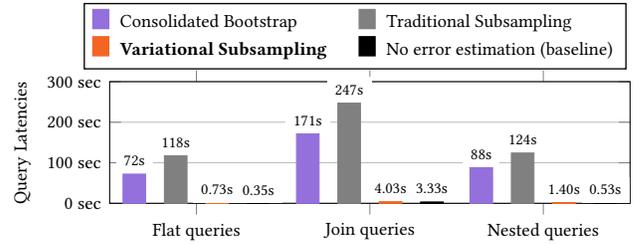

\begin{figure*}[t]
    \pgfplotsset{subsampling/.style={
        width=68mm,
        height=35mm,
        ymax=1.2,
        ymin=0,
        xmin=0.05,
        xmax=0.95,
        ylabel=Relative Error,
        xlabel near ticks,
        ylabel near ticks,
        ylabel style={align=center},
        legend style={
            at={(0.4,0.6)},anchor=south west,column sep=2pt,
            draw=black,fill=white,line width=.5pt,
            /tikz/every even column/.append style={column sep=20pt}
        },
        legend cell align={left},
        legend columns=1,
        every axis/.append style={font=\footnotesize},
        ymajorgrids,
        minor grid style=lightgray,
    }}

\begin{subfigure}[b]{0.36\textwidth}
\centering
\begin{tikzpicture}
    \begin{axis}[subsampling,
    xlabel=Selectivity,
    xtick={0, 0.1, ..., 1.1},
    ytick={0, 0.4, ..., 1.2},
    yticklabels={0\%, 0.4\%, 0.8\%, 1.2\%},
    ]


        \addplot[mark=*,mark size=1,black!20!mPurple,
        line width=0.5mm,
        ]
        table[x=x,y=y] {
        x y
        0.1 0.949
        0.2 0.632
        0.3 0.483
        0.4 0.387
        0.5 0.316
        0.6 0.258
        0.7 0.207
        0.8 0.158
        0.9 0.105
        };

        \addplot[mark=*,mark size=2,darkorange,
        only marks,
        error bars/.cd, y dir=both,y explicit,
          error mark options={
              rotate=90,
              darkorange,
              mark size=4pt,
              line width=1.5pt
          },
        ]
        table[x=x,y=y,
          y error plus expr=\thisrow{y}-\thisrow{neg},
          y error minus expr=\thisrow{pos}-\thisrow{y}
        ] {
        x y neg pos
        0.1  0.959   0.840   1.079
        0.2  0.624   0.543   0.706
        0.3  0.480   0.403   0.556
        0.4  0.385   0.331   0.438
        0.5  0.316   0.272   0.360
        0.6  0.260   0.224   0.296
        0.7  0.206   0.179   0.233
        0.8  0.158   0.135   0.182
        0.9  0.106   0.091   0.120
        };

        \addlegendentry{Groundtruth Error}
        \addlegendentry{Variational Subsampling}

    \end{axis}
\end{tikzpicture}
\vspace{-1.5mm}
\caption{Estimated error for different selectivity}
\label{fig:exp:subsampling:accuracy:a}
\end{subfigure}
\quad
\begin{subfigure}[b]{0.60\textwidth}
\centering
\begin{tikzpicture}
\begin{axis}[subsampling,
        xlabel=Sample Size (\verdict's query latency with var.~subsampling),
        width=100mm,
        ybar=3mm,
        ymax=4,
        ymin=0,
        xmin=1.5,
        xmax=4.5,
        bar width=2mm,
        xtick={2,3,4},xticklabels={100K (0.67 sec),1M (0.99 sec), 10M (5.54 sec)},
        ytick={0,1,2,3,4},yticklabels={0\%,1\%,2\%,3\%,4\%},
        legend style={
            at={(0.4,0.45)},anchor=south west,column sep=2pt,
            draw=black,fill=white,line width=.5pt,
            /tikz/every even column/.append style={column sep=10pt}
        },
        legend columns=2,
        ylabel=Relative Error,
    ]

    \addplot[fill=black!20!mPurple,draw=black!40!mPurple]
    table[x=x,y=y] {
    x y
    2 2.91
    3 0.92
    4 0.29
    };

    \addplot[fill=darkteal,draw=black!20!darkteal,
    error bars/.cd, y dir=both,y explicit,
        error mark options={
            rotate=90,
            black,
            mark size=5pt,
            line width=1.5pt
        },
    ]
    table[x=x,y=y,
        y error plus expr=\thisrow{y}-\thisrow{neg},
        y error minus expr=\thisrow{pos}-\thisrow{y}
    ] {
    x y neg pos
    2 2.88  2.66  3.08
    3 0.91  0.89  0.93
    4 0.29  0.29  0.29
    };

    \addplot[fill=gray,draw=black!20!gray,
    error bars/.cd, y dir=both,y explicit,
        error mark options={
            rotate=90,
            black,
            mark size=5pt,
            line width=1.5pt
        },
    ]
    table[x=x,y=y,
        y error plus expr=\thisrow{y}-\thisrow{neg},
        y error minus expr=\thisrow{pos}-\thisrow{y}
        ] {
    x y neg pos
    2 2.86  2.45  3.27
    3 0.91  0.80  1.02
    4 0.29  0.25  0.32
    };

    \addplot[fill=seablue,draw=black!20!seablue,
    error bars/.cd, y dir=both,y explicit,
        error mark options={
            rotate=90,
            black,
            mark size=5pt,
            line width=1.5pt
        },
    ]
    table[x=x,y=y,
        y error plus expr=\thisrow{y}-\thisrow{neg},
        y error minus expr=\thisrow{pos}-\thisrow{y}
    ] {
    x y neg pos
    2 2.87  2.49  3.26
    3 0.91  0.80  1.01
    4 0.29  0.26  0.32
    };

    \addplot[fill=darkorange,draw=black!20!darkorange,
    error bars/.cd, y dir=both,y explicit,
        error mark options={
            rotate=90,
            black,
            mark size=5pt,
            line width=1.5pt
        },
    ]
    table[x=x,y=y,
        y error plus expr=\thisrow{y}-\thisrow{neg},
        y error minus expr=\thisrow{pos}-\thisrow{y}
    ] {
    x y neg pos
    2 2.87  2.47  3.29
    3 0.91  0.80  1.02
    4 0.29  0.26  0.31
    };

    \addlegendentry{Groundtruth Error}
    \addlegendentry{CLT}
    \addlegendentry{Bootstrap}
    \addlegendentry{Subsampling}
    \addlegendentry{Var.~Subsampling (ours)}

\end{axis}
\end{tikzpicture}
\caption{\rev{C8}{Estimated error for different sample sizes
}}
\label{fig:exp:subsampling:accuracy:b}
\end{subfigure}

\caption{The accuracy of variational subsampling's error estimation (the error bars are the 5th and 95th percentiles).}
\label{fig:exp:subsampling:accuracy}
\end{figure*}
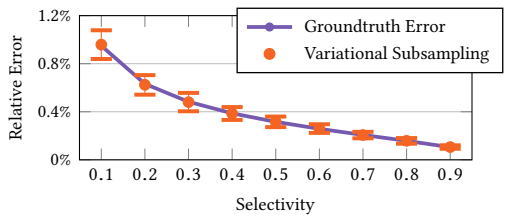
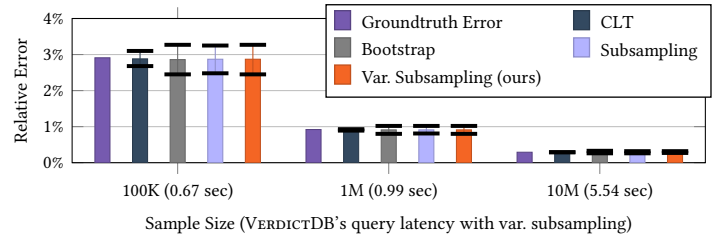

\subsection{Variational Subsampling: Correctness}
\label{sec:subsampling:error}

We study the impact of different parameters on the accuracy of variational subsampling: the query selectivity for a \texttt{count} query and the size of the sample for an \texttt{avg} query. For the latter, we also compare variational subsampling to three other methods:  central limit theorem (CLT), bootstrap, and traditional subsampling.

For this analysis, we used synthetic queries and datasets to easily control their statistical properties. The attribute values had a mean of 10.0 and a standard deviation of 10.0. To assess the quality of the error estimates, we generated 1,000 independent random samples (each sample was a subset) of the original dataset, recorded the estimated errors based on each random sample, and finally measured three statistics of the estimated errors: mean, 5th, and 95th percentiles.

First, \cref{fig:exp:subsampling:accuracy:a} depicts the estimated errors together with the groundtruth relative errors. The sample size, $n$, was 10K. The ground-truth errors were computed based on our statistical knowledge of the original data. The \emph{relative} errors decreased as the selectivity increased, since the answers to \texttt{count} queries themselves were larger with larger selectivities. Overall, variational subsampling's error estimates were within 7\% of the groundtruth. The next experiment shows that this deviation is to be expected, due to the properties of random sampling.

Second, \cref{fig:exp:subsampling:accuracy:b} compares the quality of variational subsampling's error estimation to that of other methods. Here, the ground-truth values are shown as a  reference.
When the sample size was small (i.e., 100K), resampling-based techniques were inferior to CLT since we limited the number of resamples ($b$) to 100. However, the gap reduced with larger samples.
Compared to traditional subsampling, variational subsampling was 6.5\% less accurate for $n = 100K$ (with selectivity 0.1\%); however, the difference decreased to 4.8\% for $n = 1M$ and 0\% for $n = 10M$. Processing 10M tuples---including the variational subsampling---took only 5.54 seconds.




\section{Related Work}
\label{sec:related}



\ph{Approximate Query Processing}
Sampled-based AQP received substantial attention in the research community~\cite{surajit-optimized-stratified,dynamicp-sample-selection,join_synopses,aqua2,interactive-cidr,sciborq,mozafari_eurosys2013,ganti2000icicles,hose2006distributed,kandula2016quickr,considine2004approximate,meliou2009approximating,potti2015daq,fan2015querying,mozafari_icde2010,xu2008confidence,considine2004approximate,gatterbauer2015approximate,olteanu2010approximate,arasu2004approximate,mozafari_pvldb2015_ksh,mozafari_icde2016,park2017active,mozafari_sigmod2017_dbl,he2018sigmod,kamat2014distributed}.
STRAT~\cite{surajit-optimized-stratified}, AQUA~\cite{aqua2}, and BlinkDB~\cite{mozafari_eurosys2013} have used different strategies for creating optimal stratified samples.
Online Aggregation (OLA)~\cite{online-agg,online-agg-mr2,online-agg-mr1,cosmos} continuously
refines its answers during query execution.
In the future, \verdict can also adopt some of the techniques proposed in the literature. In this work, however, we focused on variational subsampling, which enabled efficient error estimation for a wide-class of SQL queries without resorting to any tightly-integrated implementations.
%

\ph{Middleware-based Query Rewriting}
In our prior work, we have used query rewriting  to  enforce security policies transparently from the users~\cite{mozafari_usenix2017},
	or to speed up future queries by exploiting past query answers~\cite{mozafari_sigmod2017_dbl,mozafari_cidr2015}.
While Aqua~\cite{join_synopses}, IDEA \cite{galakatos2017revisiting}, Sesame \cite{kamat2018session} have also used
 query rewriting for AQP, \verdict supports a much wider range of queries (including non-PK-FK joins and nested queries),
 can work with modern distributed query engines (e.g., Hive, Spark, Impala, Redshift),
 	and does not rely on non-SQL code for sample creation.
For example, since Aqua relies on CLT-based closed-forms, it requires independent random variables, which means it can only support PK-FK joins. Also, due to Aqua's use of closed-forms, it cannot support UDAs.
\verdict has overcome this limitation with variational subsampling, which achieves generality without losing efficiency.
Furthermore, Aqua relies on the underlying engine's ability to enforce PK-FK relationships, a feature that is missing in most modern SQL-on-Hadoop engines.

\ph{Stratified Sample Construction Techniques}
BlinkDB~\cite{mozafari_eurosys2013} constructs stratified samples in two passes: one to count the size of each stratum,
	and another to perform reservoir sampling for each stratum.
Unfortunately, implementing a per-group reservoir sampling in SQL is highly complex.
For each stratum, the tuples must be separated, randomly shuffled (i.e., ordered by random integers generated on-the-fly), then
filtered using a \texttt{limit} clause. The computational cost increases linearly with the number of strata.
Quickr~\cite{kandula2016quickr} constructs stratified samples in one pass. While scanning the table, it counts the number of tuples (for each stratum) that have been read.
Based on this count, Quickr's sampler gradually reduces the sampling probability. Implementing this approach in SQL is not straightforward.

\begin{figure*}[t]
\centering

\begin{subfigure}[b]{\textwidth}
\begin{tikzpicture}
    \begin{axis}[speedups,clip=false,ymax=100]
        \addplot[fill=gray,draw=gray,point meta = explicit symbolic]
        table[meta=z] {
        x y z
        1	39.84	39.84$\times$
        2	1.00	1.00$\times$
        3	7.01	7.01$\times$
        4	44.07	44.07$\times$
        5	9.59	9.59$\times$
        6	4.98	4.98$\times$
        7	32.48	32.48$\times$
        8	1.00	1.00$\times$
        9	23.34	23.34$\times$
        10	12.13	12.13$\times$
        11	1.27	1.27$\times$
        12	14.20	14.2$\times$
        13	1.00	1.00$\times$
        14	0.00	spark-error
        15	22.00	22$\times$
        16	1.00	1.00$\times$
        17	6.46	6.46$\times$
        18	1	1.00$\times$
        19	10.9	10.92$\times$
        20	22.6	22.59$\times$
        21	12.2	12.18$\times$
        22	27.3	27.25$\times$
        23	12.8	12.82$\times$
        24	12.1	12.09$\times$
        25	9.6	9.59$\times$
        26	10.6	10.56$\times$
        27	12.3	12.33$\times$
        28	12.2	12.24$\times$
        29	11.0	10.98$\times$
        30	9.0	9$\times$
        31	0.3	0.34$\times$
        32	4.6	4.61$\times$
        33	5.4	5.4$\times$
        };
    \end{axis}
\end{tikzpicture}
\end{subfigure}

\begin{subfigure}[b]{\textwidth}
\begin{tikzpicture}
    \begin{axis}[speedups,ymax=100,clip=false]

        \addplot[fill=gray,draw=gray,point meta = explicit symbolic]
        table[meta=z] {
        x y z
        1	4.46	4.46$\times$
        2	1.00	1.00$\times$
        3	4.55	4.55$\times$
        4	31.20	31.2$\times$
        5	34.20	34.2$\times$
        6	4.12	4.12$\times$
        7	17.72	17.72$\times$
        8	1.00	1.00$\times$
        9	120.00	{}
        10	27.50	27.5$\times$
        11	9.82	9.82$\times$
        12	14.10	14.1$\times$
        13	1.00	1.00$\times$
        14	1.43	1.43$\times$
        15	13.29	13.29$\times$
        16	1.00	1.00$\times$
        17	11.40	11.4$\times$
        18	1.00	1.00$\times$
        19	12.69	12.69$\times$
        20	12.42	12.42$\times$
        21	18.12	18.12$\times$
        22	11.18	11.18$\times$
        23	10.78	10.78$\times$
        24	14.76	14.76$\times$
        25	24.48	24.48$\times$
        26	15.17	15.17$\times$
        27	25.14	25.14$\times$
        28	21.78	21.78$\times$
        29	22.23	22.23$\times$
        30	19.55	19.55$\times$
        31	2.71	2.71$\times$
        32	10.34	10.34$\times$
        33	39.84	39.84$\times$
        };

        \draw[fill=white,draw=white] (axis cs: 8.6,105) -- (axis cs: 9.4,115) -- (axis cs: 9.4,100) -- (axis cs: 8.6,90) -- cycle;
        \draw[draw=black] (axis cs: 8.6,105) -- (axis cs: 9.4,115);
        \draw[draw=black] (axis cs: 9.4,100) -- (axis cs: 8.6,90);

        \node[anchor=south west] at (axis cs: 9.2,110) {171$\times$};
    \end{axis}
\end{tikzpicture}
\end{subfigure}

\vspace{-2mm}

\caption{\rev{C6}{\verdict's speedups for Spark (top) and Impala (bottom).}}
\label{fig:exp:speedups2}
\end{figure*}
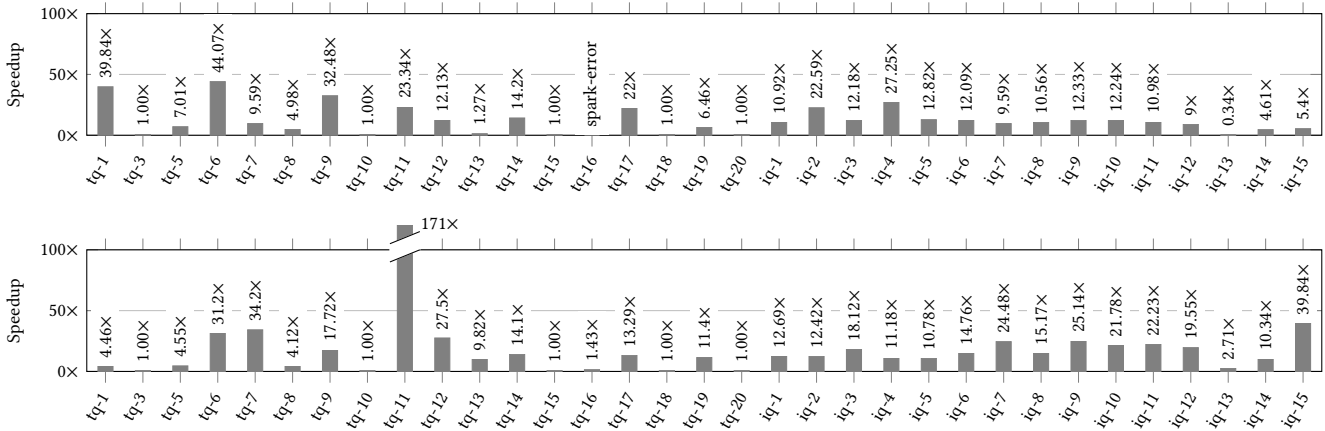

\begin{figure*}[t]
\centering
\pgfplotsset{errors/.style={
        width=180mm,
        height=30mm,
        ybar,
        ymax=10,
        ymin=0,
        xmin=0.5,
        xmax=33.5,
        bar width=2mm,
        ylabel=Relative Errors,
        xlabel near ticks,
        ylabel near ticks,
        ylabel style={align=center},
        xtick={1,2,...,33},
        xticklabels={tq-1,tq-3,tq-5,tq-6,tq-7,tq-8,tq-9,tq-10,tq-11,tq-12,tq-13,tq-14,tq-15,tq-16,tq-17,tq-18,tq-19,tq-20, iq-1,iq-2,iq-3,iq-4,iq-5,iq-6,iq-7,iq-8,iq-9,iq-10,iq-11,iq-12,iq-13,iq-14,iq-15},
        xticklabel style={rotate=60,xshift=0mm,yshift=2mm},
        ytick={0, 5, 10},
        yticklabels={0\%, 5\%, 10\%},
        legend style={
            at={(0,1.1)},anchor=south west,column sep=2pt,
            draw=black,fill=none,line width=.5pt,
            /tikz/every even column/.append style={column sep=40pt}
        },
        legend columns=2,
        every axis/.append style={font=\footnotesize},
        ymajorgrids,
        minor grid style=lightgray,
        nodes near coords,
        every node near coord/.append style={rotate=90, anchor=west},
    }}

\begin{tikzpicture}
    \begin{axis}[errors,clip=false]
        \addplot[fill=gray,draw=gray,point meta = explicit symbolic]
        table[meta=z] {
        x y z
        1   0.11  0.1\%
        2   0.00  0.0\%
        3   1.18  1.2\%
        4   0.10  0.1\%
        5   0.01  0.0\%
        6   0.17  0.2\%
        7   0.42  0.4\%
        8   0.00  0.0\%
        9   0.49  0.5\%
        10  0.98  1.0\%
        11  2.57  2.6\%
        12  0.13  0.1\%
        13  0.00  0.0\%
        14  2.53  2.5\%
        15  0.74  0.7\%
        16  0.55  0.6\%
        17  1.45  1.5\%
        18  0.32  0.3\%
        19  0   0\%
        20  0.02    0.02\%
        21  0.04    0.04\%
        22  0.001   0.001\%
        23  0.005   0.005\%
        24  0.03    0.03\%
        25  0.01    0.01\%
        26  0.03    0.03\%
        27  0.003   0.003\%
        28  0.01    0.01\%
        29  0.02    0.02\%
        30  0.06    0.06\%
        31  0.28  0.28\%
        32  0.20  0.20\%
        33  0.03  0.03\%
        };

    \end{axis}
\end{tikzpicture}
\vspace{-2mm}
\caption{Actual relative errors of the approximate answers (the associated speedups are reported in \cref{fig:exp:speedups}).}
\label{fig:exp:error1}


\end{figure*}
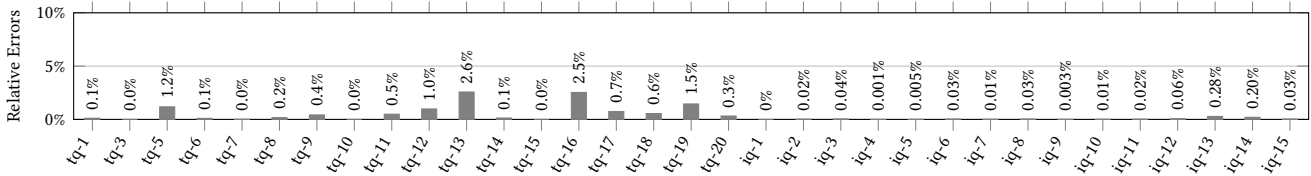

\ignore{
\begin{figure}[t]
\begin{tikzpicture}
    \begin{axis}[
        width=65mm,
        height=42mm,
        nodes near coords,
        clip=false,
        every axis/.append style={font=\footnotesize},
        ymin=0,ymax=2,xmin=0.5,xmax=5.5,
        every node near coord/.append style={rotate=60, anchor=west},
        xlabel=Sampling Ratio,
        xlabel style={yshift=-2mm},
        xtick={1,2,3,4,5},
        xticklabels={0.1\%,0.5\%,1\%,5\%,10\%},
        ytick={0,0.5,1,1.5},
        yticklabels={0,0.5,1,1.5},
        ymajorgrids,
        ]
        \addplot[draw=black,point meta = explicit symbolic,mark=*]
        table[meta=z] {
        x y z
        1 1.33183 1.33\%
        2 0.63006 0.63\%
        3 0.42 0.42\%
        4 0.18368 0.18\%
        5 0.00000 0.00\%
        };
    \end{axis}
\end{tikzpicture}
\vspace{-4mm}
\caption{Actual errors; varying sample sizes.}
\label{fig:exp:error2}
\end{figure}
}

\section{Conclusion}
\label{sec:conclusion}

In this paper, we have shown that Universal AQP (i.e., database-agnostic AQP) is a viable approach. We have proposed techniques for sample creation
and error estimation that rely solely on standard SQL queries; without making any modifications to existing databases, 
	our AQP solution can operate atop any existing SQL-based engine. Not only is our driver-level solution  comparable to fully integrated  
	AQP engines in terms of performance, in some cases it even outperforms them, thanks to its 
		novel error estimation technique, called \emph{Variational Subsampling}. 
		To the best of our knowledge, 
			we are the first to use subsampling in an AQP context.
		We also proved that, while significantly faster than traditional subsampling,
			our Variational Subsampling retains the same asymptotic properties,
			and can handle joins and complex queries.
Overall, we demonstrated that \verdict offers massive speedups (18.45$\times$ on average, and up to 171$\times$ with less than 2.6\% relative errors) to a variety of
	 popular query engines, including Impala, Spark SQL, and Amazon Redshift.


\ph{Future work}
We plan to add drivers to support additional databases (Presto, Teradata, Oracle, HP Vertica).
Our future research plans include 
%
%
(1) exploring online sampling in a middleware setting,
(2) creating a robust physical designer~\cite{mozafari_sigmod2015} to decide which samples to build,
and (3) performing a comprehensive study of
	how \verdict's approximation features affect user behavior.


\appendix

\section{Acknowledgement}
This research is in part supported by National Science Foundation through grants 1629397, 1544844, and 1553169. 
The authors are grateful to Morgan Lovay, Jeffrey Naughton, and an anonymous reviewer from Google
 for their   insightful comments.




\section{Additional Experiments}
\label{sec:more_exp}



\subsection{Actual Errors of \verdict's Answers}
\label{sec:errors}

This section reports the actual relative errors of \verdict's AQP performed in \cref{sec:speedups}. 
 \verdict's ability to estimate those actual errors are separately studied in \cref{sec:subsampling:error}.

\cref{fig:exp:error1} shows the actual relative errors for all 33 queries. 
The errors were nearly identical across different engines (module negligible differences due to the nature of random sampling);
thus, we only report the results for Impala here. 
The errors were between 0.03\%--2.57\%. The primary reason for observing different errors was due to the cardinality of the grouping attributes. For example, if there are 10$\times$ more unique values in the grouping attributes, the number of tuples averaged by AQP is reduced by 10$\times$, which in turn 
	 increases the approximation error by about $\sqrt{10}\times$ ($\approx 3.2$).

\ignore{
One limitation of the current version of \verdict is that, due to its lack of an error-bounded querying feature, we (as a user) had to make educated guesses on the I/O budget depending on the cardinality of grouping attributes. We plan to address this limitation in our future work (\cref{sec:conclusion}). \yongjoo{Want to make this part sound nicer.}
}



\subsection{Sample Preparation Time}
\label{sec:experiments:preparation}

\begin{figure}[t]
\centering
\pgfplotsset{sampling/.style={
        width=85mm,
        height=30mm,
        ybar=3mm,
        ymax=30,
        ymin=0,
        xmin=0.5,
        xmax=4.5,
        bar width=5mm,
        xlabel=Query,
        ylabel=Runtime (hours),
        xlabel near ticks,
        ylabel near ticks,
        ylabel style={align=center},
        xtick={1,2,3,4},
        xticklabels={t1,t12},
        ytick={0, 10, ..., 30},
        legend style={
            at={(1.05,1)},anchor=north west,column sep=2pt,
            draw=black,fill=none,font=\scriptsize,line width=.5pt,
            /tikz/every even column/.append style={column sep=40pt},
        },
        legend columns=1,
        every axis/.append style={font=\scriptsize},
        ymajorgrids,
        yminorgrids,
        minor grid style=lightgray,
        nodes near coords,
        every node near coord/.append style={font=\tiny},
        legend cell align={left}
    }}

\begin{tikzpicture}
    \begin{axis}[sampling,xlabel=\empty,xtick={1,2,3,4},xticklabel style={align=center},
        xticklabels={{Data transfer to\\ remote cluster},{Data transfer\\ within cluster},
        {\verdict\\ stratified\\ sampling},\snappydata\\ stratified\\ sampling},
        point meta = explicit symbolic,
        ]

        \addplot[fill=gray,draw=gray,xshift=12mm]
        table[meta=z] {
        x y z
        1 25.83 25.83h
        };

        \addplot[fill=black,draw=black,xshift=4mm]
        table[meta=z] {
        x y z
        2 7.15 7.15h
        };

        \addplot[fill=darkorange,draw=darkorange,xshift=-4mm]
        table[meta=z] {
        x y z
        3 0.59 0.59h
        };

        \addplot[fill=mPurple,draw=mPurple,xshift=-12mm]
        table[meta=z] {
        x y z
        4 0.20 0.20h
        };

    \end{axis}
\end{tikzpicture}
\vspace{-4mm}
\caption{Comparing \verdict's sampling time to other data preparation times for the 370 GB dataset.}
\label{exp:sampling:overhead}
\vspace{-4mm}
\end{figure}
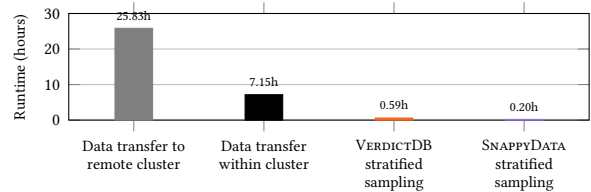

In this section, we demonstrate that \verdict's sampling preparation is sufficiently fast compared to typical tasks needed for preparing data in cluster. Since the runtime overhead of the ETL process could vary depending on the types of workloads (e.g., from simple csv parsing to entity recognition with natural language processing techniques), we compared \verdict's sampling time to the default runtime overhead that must occur: data transfer time. Also, we compare \verdict sampling time to \snappydata's sampling time.

We measured two types of data transfer overhead. The first was the data transfer to a remote cluster (i.e., scp files to an AWS instance). The second was the data transfer within a cluster (i.e., file uploads to HDFS). \cref{exp:sampling:overhead} depicts the results. \verdict's sample preparation time was much smaller compared to the other tasks. This is because sampling creation workloads are mostly read-only, which distributed storage systems (e.g., HDFS) support well. The other tasks asked heavy write loads. Even though our cluster had SSD, the runtime was still much slower.
\snappydata's sampling was faster than \verdict due to its tight integration.






\subsection{Further Study of Variational Subsampling}
\label{sec:convergence}

\begin{figure}[t]

\pgfplotsset{
  subtimeerr/.style={
    width=75mm,
    height=32mm,
    xtick={1,2,...,11},
    xticklabels={1K,2K,4K,6K,8K,10K,20K,40K,60K,80K,100K},
    xmin=5.5,
    xmax=11.5,
    ylabel=Relative Error,
    xlabel near ticks,
    ylabel near ticks,
    ylabel style={align=center},
    legend style={
        at={(0.4,0.4)},anchor=south west,column sep=2pt,
        draw=black,fill=white,line width=.5pt,
        /tikz/every even column/.append style={column sep=3pt}
    },
    legend cell align={left},
    legend columns=1,
    every axis/.append style={font=\footnotesize},
    ymajorgrids,
    minor grid style=lightgray,
}}

\begin{subfigure}[b]{\linewidth}
\hspace*{4mm}
\begin{tikzpicture}
    \begin{axis}[subtimeerr,
    xlabel=Sample Size ($n$),
    ymin=0,
    ymax=0.01,
    xticklabels={1K,2K,4K,6K,8K,10K,20K,40K,60K,80K,100K},
    ytick={0.00,0.002,0.004,0.006,0.008,0.010},
    yticklabels={0.0\%,0.2\%,0.4\%,0.6\%,0.8\%,1.0\%},
    scaled y ticks=false,
    ylabel shift=5pt,
    ]

    \addplot[
      mark=*,
      mark size=1,
      black!50!gray,
      ultra thick,
    ]
    table[x=x,y=y] {
        x y
        6   0.00348
        7   0.00225
        8   0.0017
        9   0.0013
        10  0.00109
        11  0.00097
    };

    \addplot[
      mark=*,
      mark size=1,
      mPurple,
      ultra thick,
    ]
    table[x=x,y=y] {
        x y
        6   0.00376
        7   0.0022
        8   0.00165
        9   0.00121
        10  0.001
        11  0.00088
        };

    \addplot[
      mark=*,
      mark size=1,
      darkorange,
      ultra thick,
    ]
    table[x=x,y=y] {
        x y
        6   0.00509
        7   0.00314
        8   0.0018
        9   0.00136
        10  0.00108
        11  0.00095
        };

    \addlegendentry{Bootstrap ($b$ = 1,000)}
    \addlegendentry{Subsampling ($b$ = 1,000)}
    \addlegendentry{Var.~Subsampling ($b = n^{1/2}$)}
    \end{axis}
\end{tikzpicture}
\vspace{-2mm}
\caption{Accuracy of error bound estimation}
\label{fig:exp:time_err:a}

\end{subfigure}

\vspace{2mm}

\begin{subfigure}[b]{\linewidth}
\hspace*{1mm}
\begin{tikzpicture}
    \begin{axis}[subtimeerr,
    xlabel=Sample Size ($n$),
    ymin=0.01,
    ymax=1000,
    ymode=log,
    ytick={0.01,0.1,1,10,100,1000},
    yticklabels={0.01 ms,0.1 ms,1 ms,10 ms,100 ms,1000 ms},
    scaled y ticks=false,
    ylabel=Latency,
    ]

    \addplot[
      mark=*,
      mark size=1,
      black!50!gray,
      ultra thick,
    ]
    table[x=x,y=y] {
        x y
        6   44.864
        7   85.62645
        8   157.64365
        9   232.8281
        10  304.3869
        11  283.42455
    };

    \addplot[
      mark=*,
      mark size=1,
      mPurple,
      ultra thick,
    ]
    table[x=x,y=y] {
        x y
        6   40.63755
        7   81.4604
        8   152.26045
        9   230.4664
        10  297.9938
        11  283.5998
        };

    \addplot[
      mark=*,
      mark size=1,
      darkorange,
      ultra thick,
    ]
    table[x=x,y=y] {
        x y
        6   0.0181
        7   0.022
        8   0.03815
        9   0.0439
        10  0.0651
        11  0.0682
        };
    \end{axis}
\end{tikzpicture}
\vspace{-2mm}
\caption{Latency of error bound estimation}
\label{fig:exp:time_err:b}

\end{subfigure}

\vspace{2mm}

\begin{subfigure}[b]{\linewidth}
\centering
\begin{tikzpicture}
    \begin{axis}[subtimeerr,
    xlabel=Latency,
    xmin=0.01,
    xmax=1000,
    ymin=0,
    ymax=0.006,
    xtick={0.01,0.1,1,10,100,1000},
    xticklabels={0.01 ms,0.1 ms,1 ms,10 ms,100 ms,1 sec},
    ytick={0.00,0.002,0.004,0.006},
    yticklabels={0.0\%,0.2\%,0.4\%,0.6\%},
    scaled y ticks=false,
    xmode=log,
    ]

    \addplot[
      mark=*,
      mark size=1,
      black!50!gray,
      ultra thick,
    ]
    table[x=x,y=y] {
      x y
      44.864    0.00348
      85.62645    0.00225
      157.64365   0.0017
      232.8281    0.0013
      304.3869    0.00109
      283.42455   0.00097
    };

    \addplot[
      mark=*,
      mark size=1,
      mPurple,
      ultra thick,
    ]
    table[x=x,y=y] {
      x y
      40.63755  0.00376
      81.4604 0.0022
      152.26045 0.00165
      230.4664  0.00121
      297.9938  0.001
      283.5998  0.00088
    };

    \addplot[
      mark=*,
      mark size=1,
      darkorange,
      ultra thick,
    ]
    table[x=x,y=y] {
      x y
      0.0181  0.00509
      0.022 0.00314
      0.03815 0.0018
      0.0439  0.00136
      0.0651  0.00108
      0.0682  0.00095
    };
    \end{axis}
\end{tikzpicture}
\vspace{-2mm}
\caption{Accuracy-Latency Relationship}
\label{fig:exp:time_err:c}

\end{subfigure}

\caption{Time-error tradeoff for different sample sizes ($n$).}
\label{fig:abc}
\end{figure}
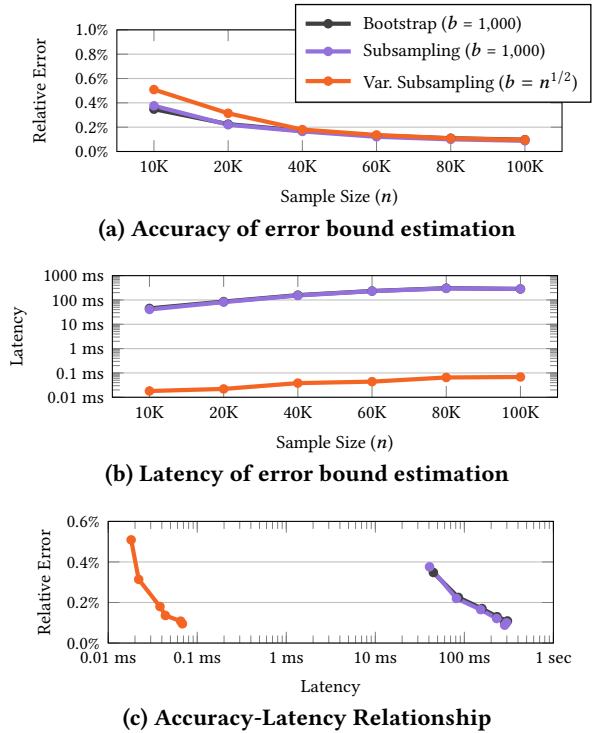

The accuracy and the convergence rate of resampling-based  techniques are typically 
studied under the assumption that the number of resamples, $b$, is very large (almost infinite).
In practice, however, the value of $b$ can considerably affect the query performance. In this section, we study 
this both empirically and theoretically.

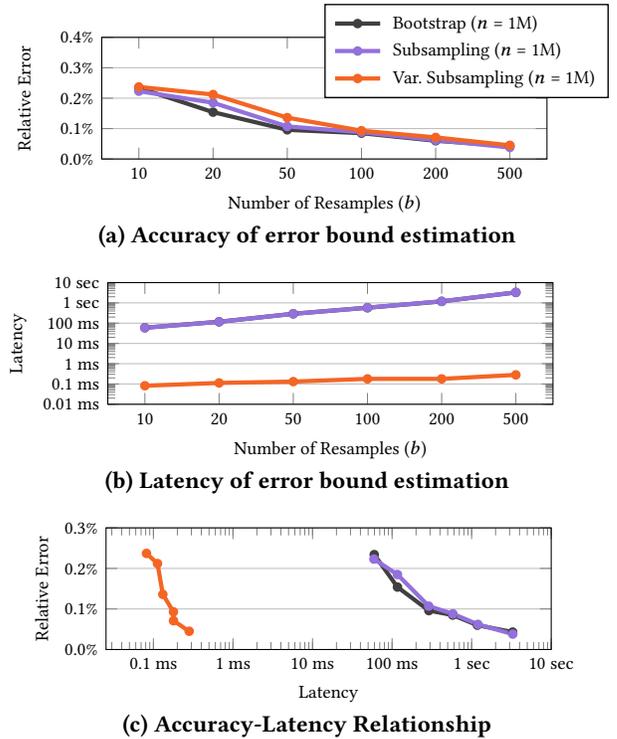
\begin{figure}[t]

\pgfplotsset{
  subtimeerr/.style={
    width=75mm,
    height=32mm,
    ymax=0.01,
    ymin=0,
    xmin=0.5,
    xmax=6.5,
    ylabel=Relative Error,
    xlabel near ticks,
    ylabel near ticks,
    ylabel style={align=center},
    legend style={
        at={(0.5,0.5)},anchor=south west,column sep=2pt,
        draw=black,fill=white,line width=.5pt,
        /tikz/every even column/.append style={column sep=3pt}
    },
    legend cell align={left},
    legend columns=1,
    every axis/.append style={font=\footnotesize},
    ymajorgrids,
    minor grid style=lightgray,
}}

\begin{subfigure}[b]{\linewidth}
\hspace*{2mm}
\begin{tikzpicture}
    \begin{axis}[subtimeerr,
    xlabel=Number of Resamples ($b$),
    ymin=0,
    ymax=0.004,
    xtick={1,2,3,4,5,6},
    xticklabels={10,20,50,100,200,500},
    ytick={0.00,0.001,0.002,0.003,0.004},
    yticklabels={0.0\%,0.1\%,0.2\%,0.3\%,0.4\%},
    scaled y ticks=false,
    ylabel shift=5pt,
    ]

    \addplot[
      mark=*,
      mark size=1,
      black!50!gray,
      ultra thick,
    ]
    table[x=x,y=y] {
      x y
      1 0.00234
      2 0.00154
      3 0.00096
      4 0.00085
      5 0.0006
      6 0.00043
    };

    \addplot[
      mark=*,
      mark size=1,
      mPurple,
      ultra thick,
    ]
    table[x=x,y=y] {
      x y
      1 0.00223
      2 0.00185
      3 0.00107
      4 0.00088
      5 0.00062
      6 0.00038
    };

    \addplot[
      mark=*,
      mark size=1,
      darkorange,
      ultra thick,
    ]
    table[x=x,y=y] {
      x y
      1 0.00237
      2 0.00212
      3 0.00136
      4 0.00093
      5 0.00071
      6 0.00045
    };

    \addlegendentry{Bootstrap ($n$ = 1M)}
    \addlegendentry{Subsampling ($n$ = 1M)}
    \addlegendentry{Var.~Subsampling ($n$ = 1M)}
    \end{axis}
\end{tikzpicture}
\vspace{-2mm}
\caption{Accuracy of error bound estimation}
\label{fig:exp:time_err2:a}

\end{subfigure}

\vspace{2mm}

\begin{subfigure}[b]{\linewidth}
\hspace*{1mm}
\begin{tikzpicture}
    \begin{axis}[subtimeerr,
    xlabel=Number of Resamples ($b$),
    ymin=0.01,
    ymax=10000,
    ymode=log,
    xtick={1,2,3,4,5,6},
    xticklabels={10,20,50,100,200,500},
    ytick={0.01,0.1,1,10,100,1000,10000},
    yticklabels={0.01 ms,0.1 ms,1 ms,10 ms,100 ms,1 sec,10 sec},
    scaled y ticks=false,
    ylabel=Latency,
    ]

    \addplot[
      mark=*,
      mark size=1,
      black!50!gray,
      ultra thick,
    ]
    table[x=x,y=y] {
      x y
      1 59.6544
      2 116.869
      3 289.8603
      4 577.283
      5 1180.7959
      6 3275.923
    };

    \addplot[
      mark=*,
      mark size=1,
      mPurple,
      ultra thick,
    ]
    table[x=x,y=y] {
      x y
      1 59.076
      2 116.7357
      3 288.8741
      4 580.8211
      5 1198.8913
      6 3287.0072
    };

    \addplot[
      mark=*,
      mark size=1,
      darkorange,
      ultra thick,
    ]
    table[x=x,y=y] {
      x y
      1 0.0819
      2 0.1126
      3 0.1314
      4 0.1787
      5 0.1789
      6 0.2817
    };
    \end{axis}
\end{tikzpicture}
\vspace{-2mm}
\caption{Latency of error bound estimation}
\label{fig:exp:time_err2:b}

\end{subfigure}

\vspace{2mm}

\begin{subfigure}[b]{\linewidth}
\centering
\begin{tikzpicture}
    \begin{axis}[subtimeerr,
    xlabel=Latency,
    xmin=0,
    xmax=10000,
    ymin=0,
    ymax=0.003,
    xtick={0.01,0.1,1,10,100,1000,10000},
    xticklabels={0.01 ms,0.1 ms,1 ms,10 ms,100 ms,1 sec,10 sec},
    ytick={0.00,0.001,0.002,0.003,0.004},
    yticklabels={0.0\%,0.1\%,0.2\%,0.3\%,0.4\%},
    scaled y ticks=false,
    xmode=log,
    ]

    \addplot[
      mark=*,
      mark size=1,
      black!50!gray,
      ultra thick,
    ]
    table[x=x,y=y] {
      x y
  		59.6544 0.00234
      116.869 0.00154
      289.8603  0.00096
      577.283 0.00085
      1180.7959 0.0006
      3275.923  0.00043
    };

    \addplot[
      mark=*,
      mark size=1,
      mPurple,
      ultra thick,
    ]
    table[x=x,y=y] {
      x y
      59.076  0.00223
      116.7357  0.00185
      288.8741  0.00107
      580.8211  0.00088
      1198.8913 0.00062
      3287.0072 0.00038
		};

		\addplot[
      mark=*,
      mark size=1,
      darkorange,
      ultra thick,
    ]
    table[x=x,y=y] {
      x y
      0.0819  0.00237
      0.1126  0.00212
      0.1314  0.00136
      0.1787  0.00093
      0.1789  0.00071
      0.2817  0.00045
		};
    \end{axis}
\end{tikzpicture}
\vspace{-2mm}
\caption{Accuracy-Latency Relationship}
\label{fig:exp:time_err2:c}
\end{subfigure}

\vspace{-2mm}
\caption{Time-error tradeoff for different numbers of resamples ($b$).}
\label{fig:exp:rev:resamples_count}
\end{figure}

\ph{Comparison Against Other Techniques}
We empirically compared variational subsampling against  bootstrap and traditional subsampling,
in terms of both accuracy and latency.
In general, the accuracy of resampling-based error estimation techniques increases  as $n$ and $b$ increase. 
To verify this, we first varied  $n$ (from 10,000 to 100,000), and measured the latency and accuracy of computing 
an  error bound with 95\% confidence. 
We measured accuracy  using the relative error with respect to the true mean. 
For instance, if the true mean was \$100.0, the estimated upper bound was \$110.1, and the true upper bound was \$110.0, then the relative error of the estimated error bound was computed as (|\$110.1 - \$110.0| / \$100.0 * 100)\% = 0.1\%. 
The number of resamples ($b$) was fixed to 1,000 for bootstrap and traditional subsampling. For variational subsampling, $b$ was set to $n^{1/2}$. 
The results of this experiments are reported in \cref{fig:exp:time_err:a,fig:exp:time_err:b,fig:exp:rev:resamples_count}.

\cref{fig:exp:time_err:a} shows that bootstrap produced more accurate error estimates than both traditional  and variational subsampling (i.e., the relative errors of the estimated error bounds were lower), but the accuracy gap reduced as $n$ increased. 
However, as shown in \cref{fig:exp:time_err:b}, variational subsampling was orders of magnitude faster than both bootstrap and traditional subsampling for the same sample size.
%
%
In \cref{fig:exp:rev:resamples_count}, we also show the relationship between the number of resamples ($b$) and the relative error of the estimated error bound. Due to the prohibitive costs of bootstrap and traditional subsampling, variational subsampling's relative errors were significantly lower given the same time budget.

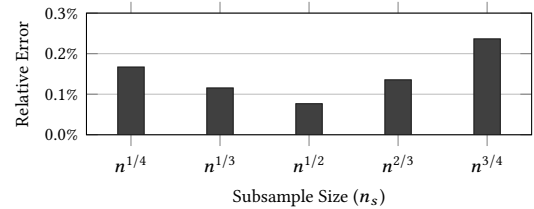
\begin{figure}[t]

\pgfplotsset{
  subtimeerr/.style={
    width=75mm,
    height=32mm,
    ybar,
    xmin=0.5,
    xmax=5.5,
    ymin=0,
    ymax=0.003,
    ylabel=Relative Error,
    xlabel near ticks,
    ylabel near ticks,
    ylabel style={align=center},
    legend style={
        at={(1.1,0.5)},anchor=south east,column sep=2pt,
        draw=black,fill=white,line width=.5pt,
        /tikz/every even column/.append style={column sep=5pt}
    },
    legend cell align={left},
    legend columns=2,
    every axis/.append style={font=\footnotesize},
    ymajorgrids,
    minor grid style=lightgray,
}}

\hspace*{-6mm}
\begin{tikzpicture}
    \begin{axis}[subtimeerr,
    xlabel=Subsample Size ($n_s$),
    xtick={1,2,...,5},
    xticklabels={$n^{1/4}$,$n^{1/3}$,$n^{1/2}$,$n^{2/3}$,$n^{3/4}$},
    ytick={0,0.001,0.002,0.003},
    yticklabels={0.0\%,0.1\%,0.2\%,0.3\%},
    scaled y ticks=false,
    ]

    \addplot[
        fill=black!50!gray,
        draw=black!80!gray,
    ]
    table[x=x,y=y] {
        x y
        1 0.001671
        2 0.001155
        3 0.000766
        4 0.001355
        5 0.002365
    };





    \end{axis}
\end{tikzpicture}


\vspace{-2mm}
\caption{The effect of the subsample size, $n_s$, on variational subsampling. The sample size was fixed, $n$=500K.}
\label{fig:exp:opt_subsample_size}
\end{figure}

\ph{Impact of Subsample Size}
When the number of resamples $b$ is small, the empirical distribution of $b$ resample-based estimates
approximates a sampling distribution.
   In this case,  an additional error term, $O(b^{-1/2})$, must be considered 
   based on
the Dvoretzky–Kiefer–Wolfowitz inequality (page 59, \cite{serfling2009approximation}).

With a finite $b$, the error of both traditional subsampling and variational subsampling
is in the order of $n_s^{-1/2} + n_s/n + b^{-1/2}$.
Since variational subsampling uses $b = n/n_s$ by default,
the error term becomes $n_s^{-1/2} + n_s/n + (n/n_s)^{-1/2}$, which can be used to 
 derive an optimal value for $n_s$.
Observe that the second term can simply be ignored since it shrinks faster than the third term. Setting the derivative of $n_s^{-1/2} + (n/n_s)^{-1/2}$  to zero produces $n_s = n^{-1/2}$.
In other words, the error expression is minimized when $n_s = n^{-1/2}$.

To empirically validate this choice, we measured the relative errors of the error bound estimates for several 
choices  of  $n_s$, namely $n^{1/4}$, $n^{1/3}$, $n^{1/2}$, $n^{2/3}$ and $n^{3/4}$. \cref{fig:exp:opt_subsample_size} shows the results. Here, the sample size, $n$, was fixed to 50,000. The results show that \verdict's default policy (i.e., $n_s = n^{1/2}$) yields the lowest errors.



\section{Proofs}
\label{sec:proofs}

In this section, we present the deferred proofs to \cref{thm:stratified}, \cref{thm:exclusive}, and \cref{thm:join}.
For each theorem, we repeat the theorem for convenience and present its proof.

\theoremstyle{theorem}
\newtheorem*{thm:stratified}{\bf\cref{thm:stratified}}
\begin{thm:stratified}
Let a sample be constructed by Bernoulli sampling from $n$ tuples with $p$ sampling probability. 
Then, the sampling probability for outputting at least $m$ tuples with probability $1 - \delta$ is
\begin{align*}
f_m(n) &= g^{-1}(m; n) \\
\text{where} \quad
g(p; n) &= \sqrt{2 n \cdot p (1 - p)} \, \erfc^{-1} \left( 2 (1 - \delta) \right) + n \, p
\end{align*}
$\erfc^{-1}$ is the inverse of the (standard) complementary error function.
\end{thm:stratified}

\begin{proof}[Proof of \cref{thm:stratified}]
Let $X$ denote the number of sampled tuples. Since each tuple is sampled independently with probability $p$ and there are $N$ such tuples, $X$ follows the Binomial distribution $B(N,\, p)$. We want $p$ to be large enough to satisfy $\Pr \left( X \ge m \right) \ge 1 - \delta$. With a standard approximation of $B(N,\, p)$ with a normal distribution $\mathcal{N}(N \cdot p, \;N \cdot p \cdot (1 - p) )$, we have
\begin{align*}
\int_{m}^{\infty} \frac{1}{\sqrt{2 \pi}}
\exp \left(
  - \frac{(x - N \cdot p)^2}{2 N \cdot p \cdot (1 - p)}
\right)
\ge 1 - \delta
\end{align*}
Then,
\begin{align*}
g(p; N) &= \sqrt{2 N \cdot p (1 - p)} \, \erfc^{-1} \left( 2 (1 - \delta) \right) + N \, p
\ge m \\
p &\ge g^{-1} (m; N)
\qedhere
\end{align*}
\end{proof}

\vspace{4mm}

\newtheorem*{thm:exclusive}{\bf\cref{thm:exclusive}}
\begin{thm:exclusive}
Let $J_n(x)$ denote the (non-degenerate) true distribution (cumulative distribution function) of the estimate based on a sample of size $n$.
Then, for any $n_s$ such that $n_s \rightarrow \infty$ and $n_s/n \rightarrow 0$ as $n \rightarrow \infty$,
\[
L_n(x) \rightarrow J_n(x)
\]
in distribution as $n \rightarrow \infty$.
\end{thm:exclusive}

\tofixnew{It is important to understand the difference between the proofs of traditional subsampling \cite{politis1994large} and variational subsampling. Traditional subsampling creates $\binom{n}{n_s}$ (= $b$) subsamples (each of size $n_s$) and computes aggregates of each subsample (i.e., $Y_1, \ldots, Y_b$). The critical part of the original proof is to show that the distribution of $(Y_1, \ldots, Y_b)$ converges to the true distribution of the aggregates of the samples of size $n_s$. In the case of traditional subsampling, $(Y_1, \ldots, Y_b)$ forms a U-statistics with a kernel of degree $n_s$. Then, according to Hoeffding's inequality for U-statistics, the distribution of aggregates converges in probability to the true distribution under the assumption that $n/n_s$ goes to infinity.}

\tofixnew{In contrast, variational subsampling relies on $b$ non-overlapping sets of random variables, where the size of the $i$-th set is $n_{s,i}$. In other words, the first subsample is a set of $X_1, \ldots, X_{n_{s,1}}$; the second subsample is a set of $X_{n_{s,1}+1}, \ldots, X_{n_{s,1}+n_{s,2}}$; and so on. Similar to the above case, let $Y_1, ..., Y_b$ denote the aggregates of subsamples. In variational subsampling, $Y_1, ..., Y_b$ are mutually independent because each of them is an aggregate of distinct iid random variables. The main part of our proof is to show that the distribution of $(Y_1, \ldots, Y_b)$ converges to the true distribution. For this, our proof employs the simple version of Hoeffding's inequality that works on independent random variables.}

\begin{proof}[Proof of \cref{thm:exclusive}]
$L_n(x)$ can be decomposed as follows.
\begin{align}
L_n(x) &= \frac{1}{b} \sum_{i=1}^b \bm{1} \left( \sqrt{n_{s,i}} ( \gs_i - \gs_0 ) \le x \right) \nonumber  \\
&= \frac{1}{b} \sum_{i=1}^b \bm{1} \left(
\sqrt{n_{s,i}} ( \gs_i - g ) + \sqrt{n_{s,i}} (g - \gs_0) \le x
 \right)
\label{eq:extended}
\end{align}
Observe that
$\frac{1}{b} \sum_{i=1}^b \bm{1} \left(
\sqrt{n_{s,i}} ( \gs_i - g ) \le x \right)$
converges to $J_n(x)$ since $n_{s,i} \rightarrow \infty$ as $n \rightarrow \infty$.
Therefore, if the second term, namely $\sqrt{n_{s,i}} (g - \gs_0)$, vanishes to 0, our theorem holds.

According to Hoeffding's inequality,
\begin{align*}
\Pr \left( | \gs_i - E \gs_i | \ge \frac{\varepsilon}{\sqrt{n_{s,i}}} \right) 
< 2 \cdot \exp ( - n \cdot \varepsilon^2 / n_{s,i} )
\end{align*}

If $n \cdot \varepsilon^2 / n_{s,i} \rightarrow \infty $  as $n \rightarrow \infty$, then
\[
\Pr \left( | \gs_i - E \gs_i | \ge \frac{\varepsilon}{\sqrt{n_{s,i}}} \right) \rightarrow 0
\]
as $n \rightarrow \infty$ for any $\varepsilon$. This means that the second term
in \cref{eq:extended} converges to 0 in probability as $n \rightarrow \infty$.

Thus, we show $n \cdot \varepsilon^2 / n_{s,i} \rightarrow \infty$ as $n \rightarrow \infty$.
Observe that $n_{s,i}$ is a binomial random variable $\mathcal{B}(n, 1 / \sqrt{n})$.
Therefore, the variance of $n_{s,i} / n$ can be expressed as
$\left( n \cdot \frac{1}{\sqrt{n}} \left( 1 - \frac{1}{\sqrt{n}} \right) \right) \Biggm/ n =
 \frac{1}{\sqrt{n}} - \frac{1}{n}$. This variance converges to 0 as $n \rightarrow \infty$. Since the variance converges to 0, the probability that $n_{s,i} / n$ is arbitrarily close to 0 is 1.0. Therefore, $n_{s,i} / n \rightarrow 0$ in probability.\footnote{``$x$ converges in probability to $y$'' means that the probability of the absolute difference between $x$ and $y$ being larger than any $\varepsilon > 0$ converges to 0.}
As stated above, this implies that the second term
in \cref{eq:extended} converges to 0 as $n \rightarrow \infty$.

Since $b \rightarrow \infty$ as $n \rightarrow \infty$,
$L_n(x)$ becomes an empirical distribution using an infinite number of samples. Thus, it
converges to the true distribution.
\end{proof}

\vspace{4mm}

\noindent
\newtheorem*{thm:join}{\bf\cref{thm:join}}
\begin{thm:join}
If there exists
\begin{align}
h(i,j) = k \quad \text{for } (i,j) \in (\mathcal{I} \times \mathcal{J})_k = \mathcal{I}_k \times \mathcal{J}_k
\end{align}
then
\begin{align}
(T \bowtie S)_{v} = \Pi_{*,\, h(i,j) \; \texttt{as}\;  \texttt{sid}} \; \left( T_{v} \; \bowtie \; S_{v} \right)
\qedhere
\end{align}
\end{thm:join}

\begin{proof}[Proof of \cref{thm:join}]
From \cref{eq:join_var1},
\begin{align*}
(T \bowtie S)_{v,k}
\; &= \;
\cup_{i \in \mathcal{I}_k} T_{v,i} \; \bowtie \; \cup_{j \in \mathcal{J}_k } S_{v,j} \\
\; &= \; 
\bigcup_{(i,j) \in (\mathcal{I} \times \mathcal{J})_k} \;
 T_{v,i} \; \bowtie \; S_{v,j}  \\
\; &= \; \sigma_{h(i,j) = k} \left(
T_{v} \; \bowtie \; S_{v}
\right)
\end{align*}
where $\sigma_{h(i,j) = k}$ is the selection operator.

Based on the above equation,
the variational table of the join, namely $(T \bowtie S)_{v}$, can be expressed as
\begin{align*}
(T \bowtie S)_{v} &= \bigcup_{k=1, \ldots, b} \; (T \bowtie S)_{v,k} \\
&= \bigcup_{k=1, \ldots, b} \;
   \sigma_{\texttt{sid} = k} \; \left(
     \Pi_{*,\, h(i,j) \text{ as } \texttt{sid}} \;
     \left(  T_{v} \; \bowtie \; S_{v}  \right)  
   \right) \\
&= \Pi_{*,\, h(i,j) \text{ as } \texttt{sid}} \; \left( T_{v} \; \bowtie \; S_{v} \right)
\qedhere
\end{align*}
\end{proof}

\vspace{4mm}

\noindent
\newtheorem*{thm:join_subsample}{\bf\cref{thm:join_subsample}}
\begin{thm:join_subsample}
Let $g(T, S)$ be an aggregate function involving two tables $T$ and $S$,
and $\gs(T_s, S_s)$ be an estimator of $g(T, S)$, where $T_s$ and $S_s$ are respective samples of $T$ and $S$.
Furthermore, let $T_{s,i}$ and $S_{s,j}$ be the $i$-th and the $j$-th subsamples of $T_s$ and $S_s$, respectively. Lastly, let $n_{s,i,j}$ denote the cardinality of the join of $T_{s,i}$ and $S_{s,j}$.
If $|T_s|/|T_{s,i}| = |S_s| / |S_{s,j}|$,
\[
L_n(x) = \frac{1}{b^2} \sum_{i=1}^{b} \sum_{j=1}^{b} 
\mathds{1} \left( \sqrt{n_{s,i,j}} \, (\gs(T_{s,i}, S_{s,j}) - \gs(T_s, S_s)) \le x \right)
\]
converges to the true distribution $J_n(x)$ of $\gs(T_s, S_s)$ as $n \rightarrow \infty$.
\end{thm:join_subsample}

\begin{proof}[Proof of \cref{thm:join_subsample}]
Define
\begin{equation}
U_n(x) = \frac{1}{b^2} \sum_{i=1}^{b} \sum_{j=1}^{b}
\mathds{1} \left( \sqrt{n_{s,i,j}} \, (\gs(T_{s,i}, S_{s,j}) - g(T, S)) \le x \right)
\label{eq:join:a}
\end{equation}
To show that $L_n(x)$ converges to the true distribution $J_n(x)$, it suffices to show that the above U-statistic, i.e., $U_n(x)$, converges to the true distribution of $\gs(T_s, S_s)$ (see the proof of theorem 2.1 in \cite{politis1994large}).
When \cref{eq:join:a} involves subsamples of a single sample, Hoeffding's inequality for U-statistics can be applied to show the convergence of $U_n(x)$ to the true distribution. However,
since \cref{eq:join:a} involves respective subsamples of $T_s$ and $S_s$,
Hoeffding's inequality is not directly applicable.

To show the convergence of $U_n(x)$ to the distribution $J_n(x)$ of $\gs(T_s, S_s)$, we employ the result on two-sample statistics \cite{lehmann1951consistency}.
This result indicates that, if the value of $U_n(x)$ does not depend on the orders of the sampled tuples, $\sqrt{n} (U_n(x) - E(U_n(x)))$ is asymptotically normally distributed (with mean zero), as $|T_s| \rightarrow \infty$.
Note that $U_n(x)$ is an unbiased estimator of the true distribution; thus, $E(U_n(x))$ is simply $J_n(x)$.

An implication of the above result is that the variance of $\sqrt{n} (U_n(x) - E(J_n(x)))$ is finite. This implies that,
for any
$x$,
 $\text{Var}(U_n(x) - J_n(x)) \rightarrow 0$ as $n \rightarrow \infty$. Thus, $U_n(x)$ converges to $J_n(x)$ as $n \rightarrow \infty$.
 \qedhere

\ignore{
\tofixnew{
To prove this theorem, we define a single-sample function $h'(C_s)$ that produces the same output as $g'(T_s, S_s)$. Note that traditional subsampling \cite{politis1994large} requires that a sample consists of i.i.d.~random variables drawn from an arbitrary sample space. All we need to do is to define the probability distribution $P$ of the random variables, which is used to construct $C_s$. One will find that with our definition of $P$, it is straightforward $h'(C_s)$ outputs the same value as $g'(T_s, S_s)$.
Also, it is straightforward that a subsample (simple random sample without replacement) $C_{s,i}$ of $C_s$ amounts to a combination of $T_{s,i}$ and $S_{s,i}$.

Without loss of generality, we assume that the dimension (i.e., the number of columns) $d_T$ of $T$ is equal to or larger than the dimension $d_S$ of $S$. The $i$-th element $c_i$ of $C_s$ (which is a random variable) is obtained as follows. With chance of $|T_s| / (|T_s| + |S_s|)$, $c_i$ is set as $(1, t_i)$ where $t_i$ is a randomly chosen tuple of $T$; with chance of $|S_s| / (|T_s| + |S_s|)$, $c_i$ is set as $(2, s_i, 0, \ldots, 0)$ where $s_i$ is a randomly chosen tuple of $S$ and 0 is padded $d_T - d_S$ times at the end (to make the size $c_i$ equal). Observe that $c_i$ for $i = 1, \ldots, n$ are i.i.d.~where $n = |T_s| + |S_s|$ is the size of $C_s$.

Next, $h'$ is defined as follows. Initially, $h'$ creates empty sets $A$ and $B$. For every $c_i$, it checks $c_i$'s first element; if the value is 1, $c_i$ is inserted into a set $A$; otherwise, $c_i$ is inserted into a set $B$. After classifying all elements, $h'$ outputs $g'(A, B)$. Observe that this classification process essentially identifies the original table to which a sampled tuple belongs.

We could use $h'(C_{s,i})$ for $i = 1, \ldots, b$ to estimate the distribution of $h'(C_s)$ (which is equal to $g'(T_s, S_s)$). Since $h'(C_{s,i}) = g'(T_{s,i}, S_{s,i})$ with our definition of $C_s$ and $h'$, we can simply use $g'(T_{s,i}, S_{s,i})$ for $i = 1, \ldots, b$ to estimate the distribution of $g'(T_s, S_s)$.
}
}

\ignore{
To prove this theorem, we will first define a single-sample function $h(C)$ that outputs the same value as $g(T,S)$. We will also define an estimator $h'(C_s)$ whose outputs follows the same distribution as $g'(T_s, S_s)$. Since $h'(C_s)$ involves a single sample ($C_s$), according to subsampling, the empirical distribution of $h'(C_{s,i})$ (where $C_{s,i}$ is the $i$-th subsample of $C_s$) converges to the true distribution of $h'(C_s)$ as $b \rightarrow \infty$ (when properly scaled as described in \cref{sec:subsampling}). Finally, we will show that the output of $g'(T_{s,i}, S_{s,i})$ follows the same distribution as $h'(C_{s,i})$; thus, the empirical distribution of $g'(T_{s,i}, S_{s,i})$ must also converge to the true distribution of $h'(C_s)$, thus to the true distribution of $g'(T_s, S_s)$.

Let $d = |T|/|S|$. We define the set $C$ as follows. The $i$-th tuple $c_i$ of $C$ is 
the concatenation of $d$ tuples of $T$ and a single tuple of $S$. That is, $c_i = (t_{(i-1) d+1}, \ldots, t_{i d}, s_i)$, where each of $t_{(i-1) d + 1}, \ldots, t_{i d}$ is a random variable that represents a randomly-chosen tuple of $T$ (by simple random sampling without replacement); $s_i$ is a random variable that represents a randomly-chosen tuple of $S$ (also by simple random sampling without replacement). Note that $c_i$ is also a random variable (that represents a vector).
The function $h$ is defined as follows.
Given $C$, the function $h$ reverses the above concatenation process to obtain $T$ and $S$ and performs an aggregation using $g(T, S)$.

To construct the sample $C_s$, the following sampling process for $C$ is defined. Given $C = (T, S)$, choose $d$ tuples from $T$ and a single tuple from $S$, ignoring the previously constructed concatenation for $C$; then concatenate those $d+1$ values to compose a single random variable of $C_s$. This process is repeated $|S_s|$ times.
Observe that the above sampling process is basically simple random sampling of the tuples of $T$ and $S$; thus, the chosen tuples follow the same distribution as $T_s$ and $S_s$.
Like $h$, $h'$ reverses the concatenation process to obtain $T_s$ and $S_s$ and performs an aggregation using $g'(T_s, S_s)$.

Finally, we show that $g'(T_{s,i}, S_{s,i})$ follows the same distribution as $h'(C_{s,i})$. For this, it suffices to show that the tuples in $C_{s,i}$ follows the same distribution as the tuples in $(T_{s,i}, S_{s,i})$.
Recall that the sample process for $C$ is basically simple random sampling of the tuples of $T$ and $S$; thus, the tuples of $C_{s,i}$ are also a simple random sample of the tuples of $T_s$ and $S_s$; thus, their distributions of the tuples in $C_{s,i}$ and the tuples of $T_{s,i}$ and $S_{s,i}$ are identical.
\qedhere
}

\end{proof}


\section{Data Appends}
\label{sec:appends}

\ph{Incremental Sample Maintenance}
All three types of samples, i.e., uniform sample, hashed sample, and stratified sample, are amenable to the data append. Observe that, for both uniform sample and hashed sample, the process is straightforward, since, given a sampling parameter $\tau$ (and additionally, a hash function for hashed sample), \verdict samples all tuples independently. A new batch of data can simply be sampled with the same $\tau$ and be inserted to existing sample tables.

For stratified samples, the sampling probabilities differ by the attribute values of $\mathcal{C}$. However, those sampling probabilities can easily be extracted from an existing stratified sample since \verdict stores those sampling probabilities in an extra column. Those ratios could then be used for sampling appended data. For the groups that did not exist in an existing sample, new sampling probabilities are generated and used.

\ph{Sample Consistency}
\tofix{We have observed users who regularly append new batches of data into an existing database table as a new partition (e.g., in Hive). In practice, these data ingestion operations are often handled by automated scripts. 
Notifying \verdict of the newly added data can be incorporated into these scripts. 
Then, \verdict can simply update its samples according to the newly-appended data. 
\verdict can also easily detect staleness of samples by checking the cardinality of the tables (when appends are the only form of updates
and the table names do not change frequently). 
Another mechanism that is more applicable to traditional database systems is through defining triggers that update the samples upon 
	insertion of new data.}


\section{Sample Planning in \verdict}
\label{sec:planning}

\verdict's sample planner selects a \emph{best} set of sample tables within an I/O budget; how we define \emph{best} is described below. The chosen sample tables are used for query processing and error estimation as described in \cref{sec:subsampling}.

In the rest of this section, we first describe the general idea of \verdict's sample planner and discuss its computational cost (\cref{sec:plan_example}). In \cref{sec:fast_planning}, we present \verdict's heuristics for lowering the computational cost.


\subsection{Sample Plans}
\label{sec:plan_example}

A \emph{sample plan} is a specification of which sample tables must be used for answering certain aggregate functions.
Specifically, a sample plan is a mapping from aggregate function(s) to a set of sample tables.
The goal of \verdict's sample planning is to find the best sample plan, i.e., the sample plan that results in the lowest approximation errors within a given I/O budget.
For this, \verdict generates many possible sample plans (called \emph{candidate plans}) and selects the one based on the criteria we will describe shortly.

Our presentation in this section will use the following example. A query includes three aggregate functions---\texttt{count(*)}, \texttt{avg(price)}, \texttt{count(distinct order\_id)}; its \texttt{from} clause has a join of the \texttt{orders} and \texttt{products} tables. We suppose a uniform random sample and a hashed sample have been built for the \texttt{orders} table, and a stratified sample and a hashed sample have been built for the \texttt{products} table.

\ph{Candidate Plans}
In the first step, \verdict's sample planner generates candidate plans. Specifically, each candidate plan is a dictionary structure in which the key is an aggregate function and the value is a set of sample tables that can be used for computing the aggregate function. 
In our example, four different combinations (i.e., $2 \times 2$) of sample tables can be used for answering each of the three aggregate functions. Thus, the total number of candidate plans is $4 \times 4 \times 4 = 64$. \Cref{tab:candidates} shows two of the candidate plans \verdict enumerates.

\begin{table}[t]
\begin{subfigure}{\columnwidth}
\centering
\small
\begin{tabular}{|p{40mm}|l|}
\hline
\textbf{Key (aggregate function)} & \textbf{Value (samples)} \\ \hline
\texttt{count(*)}                 & uniform sample of orders \\
                                  & stratified sample of products \\ \hline  
\texttt{avg(price)}                & uniform sample of orders \\
                                  & stratified sample of products \\ \hline  
\texttt{count(distinct order\_id)}  & hashed sample of orders \\
                                  & stratified sample of products \\ \hline
\end{tabular}
\vspace{1mm}
\caption{candidate plan \#1}
\label{plan1}
\end{subfigure}

\vspace{2mm}

\begin{subfigure}{\columnwidth}
\centering
\small
\begin{tabular}{|p{40mm}|l|}
\hline
\textbf{Key (aggregate function)} & \textbf{Value (samples)} \\ \hline
\texttt{count(*)}                  & hashed sample of orders \\
                                  & hashed sample of products \\ \hline
\texttt{avg(price)}                & hashed sample of orders \\
                                   & hashed sample of products \\ \hline
\texttt{count(distinct order\_id)}  & hashed sample of orders \\
                                  & hashed sample of products \\ \hline
\end{tabular}
\vspace{1mm}
\caption{candidate plan \#2}
\label{plan2}
\end{subfigure}

\vspace{2mm}
\caption{Examples of candidate plans.}
\label{tab:candidates}
\end{table}

In the second step, the sample planner examines if there exist any duplicate values (i.e., same set of sample tables) in each of the candidate plans. The existence of duplicate values indicates that two or more aggregate functions can be answered using the same set of sample tables.
Observe that the candidate plan in \cref{plan1} has a duplicate sample table (i.e., a uniform sample of \texttt{orders} and a stratified sample of \texttt{products}) for both \texttt{count(*)} and \texttt{avg(price)}. This means that the combination of those two sample tables can be used to answer both \texttt{count(*)} and \texttt{avg(price)} simultaneously. Also, the candidate plan in \cref{plan2} has the same combination, i.e., a hashed sample of \texttt{orders} and a hashed sample of \texttt{products}, for all three aggregates. This means that all three aggregate functions can be answered using the same set of those sample tables.

To optimize query processing, \verdict's sample planner consolidates those candidate plans. The consolidation merges the aggregate functions that share the same set of sample tables. After the consolidation, the key of each candidate plan is now a list of aggregate functions. The value is still a set of sample tables. The examples of the consolidate sample plans are presented in \cref{tab:consolidated_plan}.

\begin{table}[t]
\begin{subfigure}{\columnwidth}
\centering
\small
\begin{tabular}{|p{40mm}|l|}
\hline
\textbf{Key (aggregate function)} & \textbf{Value (samples)} \\ \hline
\texttt{count(*)}, \texttt{avg(price)}     & uniform sample of orders \\
										   & stratified sample of products \\ \hline  
\texttt{count(distinct order\_id)} & hashed sample of orders \\
								   & stratified sample of products \\ \hline
\end{tabular}
\vspace{1mm}
\caption{consolidated plan \#1}
\label{cplan1}
\end{subfigure}

\vspace{2mm}

\begin{subfigure}{\columnwidth}
\centering
\small
\begin{tabular}{|p{40mm}|l|}
\hline
\textbf{Key (aggregate function)} & \textbf{Value (samples)} \\ \hline
\texttt{count(*)}, \texttt{avg(price)}, & hashed sample of orders \\
\texttt{count(distinct order\_id)}  & hashed sample of products \\ \hline
\end{tabular}
\vspace{1mm}
\caption{consolidated plan \#2}
\label{cplan2}
\end{subfigure}

\vspace{2mm}
\caption{Examples of consolidated plans.}
\label{tab:consolidated_plan}
\end{table}

\ph{Selecting a Plan}
Here, we describe how \verdict selects the best plan among those consolidated plans. \verdict's plan selection relies on two criteria: \emph{scores} and \emph{costs}.
\verdict assigns a pair of score (which we describe below) and I/O cost to each of those consolidated candidate plans. Then, \verdict selects the plan with the highest score but within an I/O budget.

First, the I/O cost of each candidate plan is set as the total number of the tuples of the sample tables in the candidate plan. In \cref{cplan1}, the I/O cost is the number of the tuples of (1) the uniform sample of \texttt{orders}, (2) the stratified sample of \texttt{products} (count this twice), and (3) the hashed sample of \texttt{orders}.
In \cref{cplan2}, the I/O cost is the number of the tuples in the hashed sample of \texttt{orders} and the hashed sample of \texttt{products}. Only the plans whose I/O costs are within the I/O budget are eligible for a final plan. If there exists no such plan, \verdict uses the original tables (thus, no AQP).

Second, the score of a candidate plan is the square root of an effective sampling ratio multiplied by some advantage factors. We use the square root to accommodate the fact that the expected errors of mean-like statistics decrease as a factor of the square root of the sample size. In the case where a candidate plan may include more than one set of sample tables (as in \cref{cplan1}), we average those sampling ratios for the score. Note that the effective sampling ratio is different from the number of the tuples of sample tables. One important exception is when two hashed samples are equi-joined on their column sets. In this case, the sampling ratio of the joined table is equal to the smaller value between the sampling ratios of those joined hashed samples. An advantage factor is given to the case in which a stratified sample is used appropriately for a \texttt{group-by} query. More specifically, the column set on which the stratified sample is built should be a superset of the grouping attributes.

The number of candidate plans can be large when multiple sample tables have been built for each table, and two or more tables are joined in a query. That is, the number of the candidate plans increases exponentially as the number of tables that appear in a query increases. We describe how \verdict addresses this in the following subsection.

\subsection{Heuristic Sample Plans}
\label{sec:fast_planning}

To address the prohibitive computational costs of exhaustive enumerations, \verdict's sample planner employs a heuristic approach. Note that the large computational costs of the simple approach stem from the joins. The strategy of \verdict's heuristic approach is to, whenever a join occurs, early-prune the sample tables that are unlikely to be included in the final sample plan.

There are two types of cases where early pruning is tolerable. The first case is when the size of a sample table is too large. Then, any sample plan including the sample table will go beyond the I/O budget. Therefore, we can safely ignore them. The second case is when the size of a sample table is too small. The sample plans including those too small sample tables will be assigned lower scores (due to low sampling ratio); thus, they are unlikely to be chosen.

Based on these observations, \verdict's sample planner joins only the $k$ best sample tables at each point where a join occurs. The value of $k$ is configurable. A larger $k$ preserves a greater number of sample tables; thus, it is more conservative. On the other hand, a small $k$ value results in higher efficiency. The default value for $k$ is 10.

\begin{figure}[t]
\centering

\begin{tikzpicture}

\node (g) [align=left,font=\small,draw=black,dotted,fill=white] at (0,1.8) {
  $\prescript{}{\texttt{city}}{G}_{\texttt{count(*)}}$
   \texttt{orders} $\bowtie$ \texttt{order\_products}: [\\
   \checkmark \; 0.01\% irregular (uniform-stratified),\\
   \checkmark \; 1\% universe on \texttt{order\_id}
]};

\node (j) [align=left,font=\small,draw=black,dotted,fill=white] at (0,0) {
  \texttt{orders} $\bowtie$ \texttt{order\_products}: [\\
  \checkmark \; 0.01\% irregular (uniform-stratified),\\
  \hspace*{3mm} 0.01\% irregular (uniform-universe),\\
  \hspace*{3mm} 0.01\% irregular (universe-stratified),\\
  \checkmark \; 1\% universe on \texttt{order\_id}
]};

\node (a) [align=left,font=\small,draw=black,dotted,anchor=north] at (-2.1,-1.3) {
  \texttt{orders}: [\\
  \checkmark \; 1\% uniform,\\
  \checkmark \; 1\% universe on \texttt{order\_id}]
};

\node (b) [align=left,font=\small,draw=black,dotted,anchor=north] at (2.1,-1.3) {
  \texttt{order\_products}: [\\
  \checkmark \; 1\% stratified,\\
  \checkmark \; 1\% universe on \texttt{order\_id}]
};

\draw [thick,->] ($(a.north)+(0,0.0)$) -- (j);
\draw [thick,->] ($(b.north)+(0,0.0)$) -- (j);
\draw [thick,->] (j) -- (g);

\end{tikzpicture}

\caption{\verdict's heuristic approach to choosing a best set of sample tables. In this example, two best sample tables at each level (with \checkmark) are pushed up.}
\label{fig:join_tree}
\end{figure}
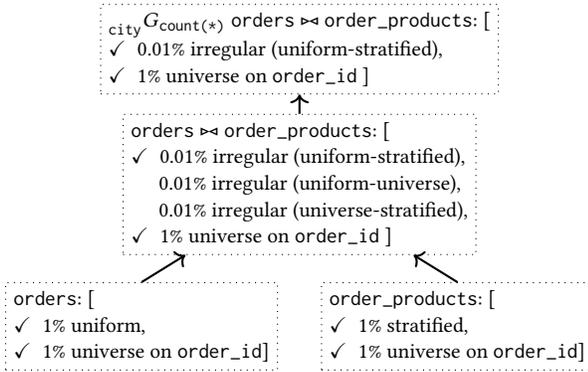

\Cref{fig:join_tree} depicts an example, in which \verdict generates candidate plans for a \texttt{count(*)} aggregate of \texttt{orders} $\bowtie$ \texttt{products} when grouped by the \texttt{city} column. $k$ is set to 2 in this example. At the bottom level, two sample tables for each of the \texttt{orders} and \texttt{products} tables are pushed up to be joined. There are four possible candidates for the joined table ($2 \times 2$). Among them, 1\% universe sample is first chosen due to its large sampling probability. Then, one of the other three 0.01\% sample tables are chosen randomly. At top level, the aggregation operation does not increase the number of candidate plans; thus, both two sample tables are included in candidates plans.

When a submitted query includes more than one aggregate function, this process repeats for every aggregate function. The rest of the process is as described in \cref{sec:plan_example}. Although we used a simple example for clear presentation, the same principle is applied to nested queries to generate multiple candidate plans and choose the best plan among them.

\section{Default Sampling Policy}
\label{sec:default_sampling}

When a user asks \verdict to create sample tables for an original table $T$ without explicitly specifying what types of samples to create, \verdict, by default, examines the cardinalities (i.e., the number of unique attribute values) in the columns of $T$ and determines the samples to build based on those cardinalities. This is based on two observations. First, stratified sample on $\mathcal{C}$ whose cardinality is too large won't benefit from sampling, since most of the tuples will be included in the sample. Second, universe sample on $\mathcal{C}$ whose cardinality is too low won't benefit, since it's mainly useful for joining two large fact tables~\cite{kandula2016quickr}.

Concretely, \verdict's default policy is as follows:
\begin{enumerate}[1.,nolistsep,noitemsep,leftmargin=10pt]
\item A sampling parameter $\tau$ is set as 10M / $|T|$.
\item \verdict always creates a uniform sample of $T$.
\item For each $\mathcal{C}_i$ of the top 10 columns (in descending  order of cardinality) whose cardinalities are larger than 1\% of $|T|$, \verdict creates a hashed sample on $\mathcal{C}_i$.
\item For each $\mathcal{C}_i$ of the top 10 columns (in ascending order of cardinality) whose cardinalities are smaller than 1\% of $|T|$, \verdict creates a stratified sample on $\mathcal{C}_i$.
\end{enumerate}

\section{Example Query Rewriting}
\label{sec:rewriting}

\tofix{In this section, we provide \verdict's query rewriting example for a simple \texttt{groupby} query.
Given the below input query:}
\begin{lstlisting}[
    basicstyle=\footnotesize\ttfamily,
    xleftmargin=5pt,
    caption={Input Query},captionpos=b,
    label={fig:rewriting:input}
]
select l_returnflag, count(*) as cc
from lineitem
group by l_returnflag;
\end{lstlisting}

\vspace{2mm}

\noindent \tofix{\verdict rewrites the above query as follows.}
\begin{lstlisting}[
    basicstyle=\footnotesize\ttfamily,
    xleftmargin=5pt,
    caption={Rewritten Query},captionpos=b,
    label={fig:rewriting:output}
]
select vt1.`l_returnflag` AS `l_returnflag`,
  round(sum((vt1.`cc` * vt1.`sub_size`))
    / sum(vt1.`sub_size`)) AS `cc`,
  (stddev(vt1.`count_order`) * sqrt(avg(vt1.`sub_size`)))
    / sqrt(sum(vt1.`sub_size`)) AS `cc_err`
from (
  select
    vt0.`l_returnflag` AS `l_returnflag`,
    ((sum((1.0 / vt0.`sampling_prob`)) /
      count(*)) * sum(count(*))
        OVER (partition BY vt0.`l_returnflag`))
    AS `cc`,
    vt0.`sid` AS `sid`,
    count(*) AS `sub_size`,
  from lineitem_sample vt0
  GROUP BY vt0.`l_returnflag`, vt0.`sid`) AS vt1
GROUP BY vt1.`l_returnflag`;
\end{lstlisting}

\tofix{The inner query computes an unbiased estimate for every subsample, which are weighted-averaged by an outer query for a final aggregate answer and an error estimation. The value in \texttt{cc\_err} contains standard deviations, which are used for computing confidence intervals for every row.}





%
%

\normalsize
\bibliographystyle{abbrv}
\bibliography{biblio/verdict,biblio/approximate,biblio/mozafari,biblio/robust,biblio/ldb}  
\normalsize

\end{document}